\documentclass[12pt]{iopart}  

\expandafter\let\csname equation*\endcsname\relax
\expandafter\let\csname endequation*\endcsname\relax
\usepackage{amsmath}
\usepackage{amsthm}
\usepackage{amssymb}
\usepackage{hyperref}
\usepackage{graphicx,bm}

\usepackage{suffix}
\usepackage{mathtools}
\usepackage{color}

\renewcommand{\O}[1]{\mathrm{O}\lb#1\rb}

\newcommand{\p}{\partial}

\newcommand{\lb}{\left(}
\newcommand{\rb}{\right)}
\newcommand{\lsb}{\left[}
\newcommand{\rsb}{\right]}

\begin{document}

\title{Logarithmic scaling of the collapse
in the critical Keller-Segel equation}

\author{Sergey A. Dyachenko, Pavel M. Lushnikov and Natalia Vladimirova}

\address{Department of Mathematics and Statistics,
University of New Mexico, Albuquerque, NM 87131, USA
}

\date{\today}

\begin{abstract}
A reduced Keller-Segel equation (RKSE) is a parabolic-elliptic system
of partial differential equations which describes bacterial
aggregation and the collapse of a self-gravitating gas of brownian
particles.  We consider RKSE in two dimensions,
 where solution has a critical collapse (blow-up) if the
total number of bacteria exceeds a critical value.  We study
the self-similar solutions of RKSE near the blow-up point.  Near
the collapse time, $t=t_c$,
the critical collapse is characterized by the $L\propto (t_c-t)^{1/2}$ scaling
law with logarithmic modification, where $L$ is the spatial width of  collapsing solution.  We develop an asymptotic
perturbation theory for these modifications and show that the
resulting scaling agrees well with numerical simulations.  The
quantitative comparison of the theory and simulations requires to take
into account several terms of the perturbation series.
\end{abstract}





\maketitle

\section{Introduction}

In this paper we consider a reduced Keller-Segel equation (RKSE)
\begin{align}
\begin{split}
& \p_t \rho=\Delta\rho-\nabla\cdot(\rho\nabla c),\\
&\Delta c=-\rho,
\end{split}
\label{pottscontinuousKellerSegelreduced}
\end{align}
which is the parabolic-elliptic system of partial differential
equations for two scalar functions, $\rho=\rho({\bf r},t)$ and
$c=c({\bf r},t)$.  Here ${\bf r}\in\Omega \subseteq \mathbb{R}^D$ is
the spatial coordinate in dimension $D$ and $t$ is the time. We assume
that either $\Omega=\mathbb{R}^D$ or $\Omega$ is a bounded domain.
For $\Omega=\mathbb{R}^D$, we also assume that both $\rho$ and $c$
decay to zero as $|{\bf r}|\to \infty$.  In the bounded domain case,
we assume the zero flux condition for both $\rho$ and $c$ through the
boundary $\partial \Omega$.

RKSE is the reduction of the well-known Keller-Segel model (also
sometimes called Patlak-Keller-Segel model).
See e.g.
\cite{Patlak1953,KellerSegel1970,Alt1980,HerreroVelazquezMathAnn1996,BrennerLevitovBudrene1998,BrennerConstantinKadanoff1999,Ben-JacobAdvPhys2000,BettertonBrennerPRE2001,VelazquezSIAMJApplMath2002,HillenOthmer2002,SireChavanis2002,ErbanOthmer2005,NewmanGrima2004,LushnikovPhysLettA2010,DejakLushnikovOvchinnikovSigalPhysD2012,DLS2009}
and references therein.  The Keller-Segel model was derived for the macroscopically averaged dynamics of bacteria  and biological cells.
Below
we refer to bacteria and cell as synonyms. Bacteria often
communicate through chemotaxis, when bacteria both secrete a substance
called chemoattractant and move along the gradient of chemoattractant.
The macroscopically averaged dynamics of bacteria
is described by the bacterial density  $\rho({\bf r},t)$ and the chemoattractant
concentration $c({\bf r},t)$.
Bacteria are self-propelled and, without the chemotactic
clue, the center of mass of each bacteria typically experiences a
random walk. The random walk is described by the first term
(diffusion) on the right-hand side (rhs) of the first equation in
\eqref{pottscontinuousKellerSegelreduced}.  The diffusion of
chemoattractant is described by the Laplacian term in the second equation.
The term on the rhs of the second equation
corresponds to the production rate of chemoattractant by the bacteria,
which is proportional to the bacterial density.  The second term on
the rhs of the first equation
characterizes the motion of
bacteria towards large values of~$c$.  The motion of bacterial colonies is
thus determined by competition between random-walk-based diffusion
and chemotaxis-based attraction.  For the convenience of the readers,
we provide a more extensive description of the Keller-Segel model and
the derivation of RKSE in~\ref{appendix:KellerSegelModel}.

Equation~\eqref{pottscontinuousKellerSegelreduced} also describes the
dynamics of a gas of self-gravitating Brownian particles, which has
applications in astrophysics including the problem of stellar
collapse~\cite{HeNeVe1997,Wo,SireChavanis2002,ChavanisSirePRE2011a}.
In this case, the second equation
in~\eqref{pottscontinuousKellerSegelreduced} is the Poisson equation
for the gravitational potential, $-c$, while $\rho$ is the gas
density.  (All units are dimensionless). The first equation
in~\eqref{pottscontinuousKellerSegelreduced} is a Smoluchowski
equation for~$\rho$.
Below we refer to $\rho$ and $c$ as the density of bacteria and the
concentration of chemoattractant, respectively, but all results below
are equally true for the gravitational collapse of a gas of
self-gravitating Brownian particles.

A solution of RKSE in dimension one (D=1) is global (in time).
For $D\ge 2$ (e.g. in dimensions two and three ) a finite time
singularity occurs~\cite{BrennerConstantinKadanoff1999} provided
the initial condition is large enough. E.g. for dimension two (D=2), a finite
time singularity occurs for $N>8\pi$, where $N=\int \rho \, d{\bf r}$
is the total number of bacteria (in rescaled units)
\cite{BilerWoyczinski1998,LushnikovPhysLettA2010}.  Below we focus on unbounded domains $\Omega=\mathbb{R}^D$.

The formation of singularity in a finite time (blow up) is a quite
general phenomenon observed in many nonlinear systems including
self-focusing in nonlinear optics, plasmas, hydrodynamics, and
collapse of Bose-Einstein condensate
\cite{KuznetsovZakharov2007}. Blow up is often accompanied by a
dramatic contraction of the spatial extent of solution, which is then
called by collapse \cite{KuznetsovZakharov2007}. Collapse typically occurs
when there are (i) self-attraction in nonlinear systems and,
(ii) a conserved quantity, such as the spatial norm (e.g., $L_2$ or
$L_1$ norm) of the solution.   Such systems are often
described by the nonlinear Schr\"odinger equation (NLSE)
\cite{Sulem1999}:
\begin{equation}
i\partial_t \psi+ \nabla^2 \psi +  |\psi|^{2} \psi =0.
\label{nls1kellersegel}
\end{equation}
 for complex
variable $\psi({\bf r},t)$. NLSE conserves the integral $P=\int
|\psi|^2 d{\bf r}$ and supports collapse for $D\ge 2$.  Similarly, RKSE describes the attraction between
brownian particles and conserves
 $L_1$ norm of $\rho$ as well as RKSE admits collapse for $D\ge 2.$

A collapse in RKSE corresponds to the aggregation of bacterial
colonies in biological applications and gravitational collapse for
self-gravitating Brownian particles.  Aggregation is a first step to a
formation of multicellular organisms and quite important in biological
applications \cite{Ben-JacobAdvPhys2000}.
E.g., the evolution of a low-density {\it
  Escherichia coli} bacteria colony in a petri dish is about one
day~\cite{BrennerLevitovBudrene1998}. However, if the bacterial density is
locally high then  bacteria aggregate on a timescale of
several minutes~\cite{BrennerLevitovBudrene1998}. Thus the aggregation has
an explosive character (see more details on that in Appendix \ref{appendix:KellerSegelModel}).
Near singularity the
Keller-Segel model is not applicable when typical distance between
bacteria is about or below the size of bacteria.  In that regime a modification of
the Keller-Segel model was derived from microscopic stochastic
dynamics of bacteria \cite{AlberChenGlimmLushnikov2006,AlberChenLushnikovNewman2007,%
LushnikovChenalberPRE2008}.  That modified model prevents collapse due to excluded volume
constraint (different bacteria cannot occupy the same volume).
Here however the original RKSE without regularization is considered.

Collapses in NLSE
and RKSE have much common, as detailed in
Ref.~\cite{LushnikovPhysLettA2010}.  E.g., the number of particles $P$ in NLSE has a similar meaning to the number of bacteria $N$ in RKSE. One can also
recall that $|\psi|^2$ is the probability density in quantum mechanics.
In
two dimensions ($D=2$), the critical number of particles, $P_c=11.70\ldots$ (for NLSE), or the
critical number of bacteria, $N_c=8\pi$ (for RKSE), determine the boundary between
collapsing and noncollapsing regimes in both systems \cite{BrennerConstantinKadanoff1999,ChiaoGarmireTownesPRL1964,VlasovPetrishchevTalanovRdiofiz1971,ZakharovJETP1972,ZakharovKuznetsovJETP1986,LushnikovJETPLett1995,SulemSulem1999}.
Collapse in the critical dimension $D=2$ is strong for both RKSE and
NLSE, which means that a finite number of bacteria (particles) is trapped within
the collapsing spatial region.  For the supercritical case ($D>2$),
collapse in both RKSE and NLSE is weak. Weak collapse implies that the
collapse is so fast that particles (bacteria) cannot keep up with the
collapse rate. Then a vanishing
number of bacteria (particles) are trapped inside the collapsing
region in the limit $t\to t_c$.

\subsection{Summary of results}

In this paper, we focus on the 2D self-similar solution of RKSE,
Eq.~\eqref{pottscontinuousKellerSegelreduced}.  We assume that the
spatial location of the collapse is ${\bf r}=0$.  Near $t_c$, in the
neighborhood of the collapse, the solution has the following radially
symmetric form:
\begin{equation}\label{pselfsimilarfull}
\begin{split}
  \rho&=\frac{1}{L(t)^{2}}\frac{8}{(1+y^2)^2},   \\
  c&=-2 \ln (1+y^2),   \\
 y&=\frac{r}{L(t)},  \quad r:={\bf r}, \\
 L(t)&\to 0 \quad \mbox{for} \quad t\to t_c.
\end{split}
 \end{equation}
Here, $L(t)$ is the time-dependent spatial width of solution.  We also refer to $L(t)$ as the collapse width.  (We
sometimes omit the argument of~$L$ for brevity.)  The self-similar
form \eqref{pselfsimilarfull} is valid in the limit $t\to t_c$ in the
small spatial neighborhood of the collapse point.  This local
applicability of the self-similar solution is typical for collapses in
numerous nonlinear systems \cite{KuznetsovZakharov2007}.

A number of different scalings for $L(t)$ have been proposed. First is the scaling,
\begin{equation}\label{Ltexplog1}
 L(t)=c\sqrt{t_c-t} e^{-\sqrt{-\frac{\ln(t_c-t)}{2}}} [-\ln{(t_c-t)}]^{(1/4)(-\ln{(t_c-t)})^{-1/2}},
 \end{equation}
where $c$ is an unknown constant. This scaling was derived in
Ref.~\cite{HerreroVelazquezMathAnn1996} using formal matched
asymptotic expansion of RKSE near \eqref{pselfsimilarfull}.

The second scaling,
\begin{equation}\label{Ltexplog2}
 L(t)=2e^{-\frac{2+\gamma}{2}}\sqrt{t_c-t}e^{-\sqrt{-\frac{\ln(t_c-t)}{2}}},
 \end{equation}
was derived in Refs.~\cite{VelazquezSIAMJApplMath2002}
and~\cite{LushnikovPhysLettA2010}.  Here $\gamma=0.577216\ldots$ is
the Euler constant.  In Ref.~\cite{VelazquezSIAMJApplMath2002}, the
formal matched asymptotic expansion of RKSE was used.
The approach in Ref.~\cite{LushnikovPhysLettA2010} was based on the expansion of the
perturbation around the collapsing solution~\eqref{pselfsimilarfull}
in terms of the eigenfunctions of the linearization operator.
Refs.~\cite{VelazquezSIAMJApplMath2002}
and~\cite{LushnikovPhysLettA2010} give different estimates of errors.

The third scaling,
\begin{equation}\label{Ltexplog3}
 L(t)=c\sqrt{t_c-t}e^{-\frac{1}{2}\sqrt{-\frac{\ln(t_c-t)\ln{[-\ln{(t_c-t)}]}}{2}}},
 \end{equation}
where $c$ is an unknown constant, was obtained in
Ref.~\cite{SireChavanis2002} by somewhat heuristic arguments.  (Also see
Ref.~\cite{SireChavanis2008} for more discussion.)

The scaling laws~\eqref{Ltexplog1}--\eqref{Ltexplog3} share two main
features: (i) the leading order square-root dependence, $L(t)\propto
\sqrt{t_c-t}$, and (ii) the logarithmic-type modifications of $L(t)$.
These modifications are necessary for the building the theory of
the collapse
in the critical dimension (2D).  Both features are strikingly similar
to the critical collapse in the 2D NLSE
\cite{ChiaoGarmireTownesPRL1964,VlasovPetrishchevTalanovRdiofiz1971,ZakharovJETP1972,SulemSulem1999,FraimanJETP1985,LeMesurierPapanicolaouSulemSulemPhysicaD1988,LandmanPapanicolaoSulemSulemPRA1988,DyachenkoNewellPushkarevZakharovPhysicaD1992,FibichPapanicolaouSIAMJApplMath1999,merleraphael2006}.
Earlier simulations~\cite{BettertonBrennerPRE2001,SireChavanis2002}
show that corrections to the leading order scaling $L(t)\propto
\sqrt{t_c-t}$ are necessary, but fail to determine the form of the
corrections.

In this paper we go much beyond the accuracy of scaling laws
\eqref{Ltexplog1}--\eqref{Ltexplog3}.  We derive a new scaling law that
agrees with the direct simulations of RKSE.  There are three main
results in this paper.

Our {\it first main result} is that~$L(t)$ is determined by the solution of the
following ordinary differential equation (ODE):
\begin{equation} \label{ataufullnextorder}
\begin{split}
& \frac{\partial_\tau a}{a^2}=-\frac{2}{\ln{\frac{1}{a}}}
+\frac{M}{(\ln{\frac{1}{a}})^2}+\frac{b_0}{(\ln{\frac{1}{a}})^3}+O\left (\frac{1}{(\ln{\frac{1}{a}})^4}\right ),  \\
& M=-2-2\gamma+2\ln{2}, \\
&b_0=\frac{\pi^2}{3} - 2 \ln^2{2} + 4\ln{2} +
 \gamma (-4 - 2 \gamma + 4\ln{2}).
\end{split}
\end{equation}
The adiabatically  slow quantity
\begin{equation}\label{adef1}
a = -L(t)\partial_t L(t),
\end{equation}
evolves over a new time scale described by a new variable, $\tau$, defined as
 \begin{equation}\label{taudef1}
 \tau=\int^t_{0} \frac{dt'}{L(t')^2}.
\end{equation}
Here and below, the notation $f(x)=O(x)$ means that there exists a
positive constant $c$ such that $|f|\le c|x|$ as $x\to 0$.  It follows
from \eqref{taudef1} that $\tau\to \infty$ as $t \to t_c$, so that
$\tau(t)$ maps the collapse time $t=t_c$ into $\tau=\infty$ in full
analogy with the ``lens transform'' of NLSE~\cite{TalanovJETPLett1970,KuznetsovTuritsynPhysLettA1985,FibichPapanicolaouSIAMJApplMath1999}.  The decrease of $L \to 0$ as $t \to t_c$ implies that $a>0$,
and the logarithmic modification of $L(t)\propto \sqrt{t_c-t}$
scaling results in $a\to 0$ as $t \to t_c$.  The logarithmic
modification also makes~$a$ a slow function of
$(t_c-t)^{1/2}$, compared with~$L$.  These scalings, as well as the
definition of~$a$, are in qualitative analogy with the scaling for NLSE collapse.

\begin{figure}
\begin{center}
\includegraphics[width = 2.98 in]{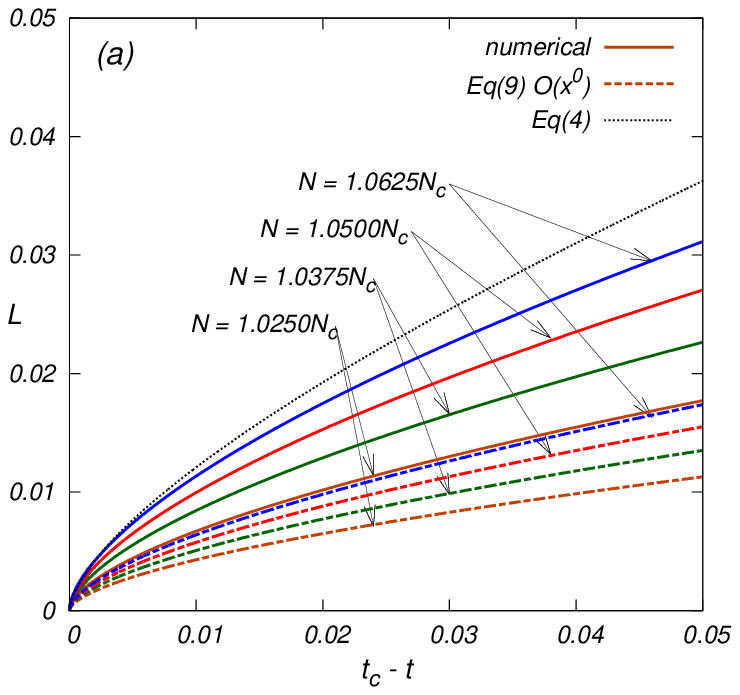}
\includegraphics[width = 2.98 in]{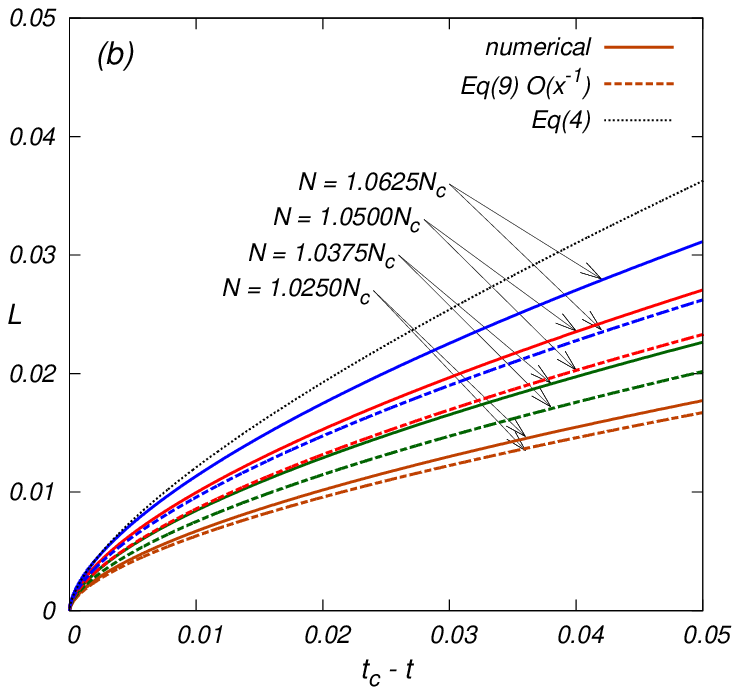}
\includegraphics[width = 2.98 in]{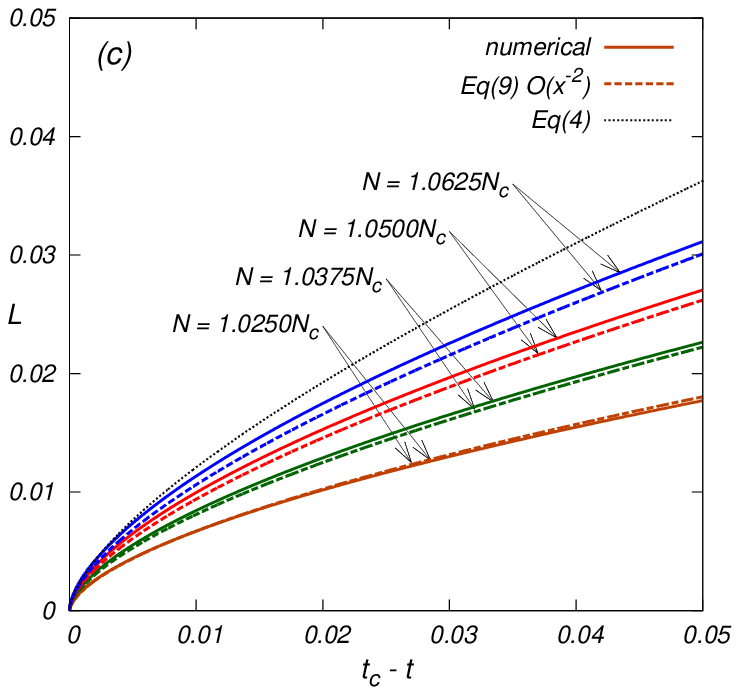}
\end{center}
\caption{Dependence $L(t)$ obtained from the numerical simulations
  of RKSE (solid lines) is compared to the
  scaling~\eqref{Ltexplog2} (dotted line) and to the
  scaling~\eqref{scaling5} (dashed-dotted lines).  The lines of different
  colors correspond to different initial conditions (different
  values of $N$).
  Different panels show the different orders
  of the scaling in the exponent of the first equation of~\eqref{scaling5}: (a)~the terms up to $O(x^{0})$ are
  taken into account; (b)~the terms up to $O(x^{-1})$ are taken into
  account; (c)~the terms up to $O(x^{-2})$, i.e., all terms except the error
  term $O(\ldots)$, are taken into account.  Convergence of the
  analytical results to the numerical results with increase of the order
  in inverse power of $x$ is clearly seen in (a)-(c).  The relative
  difference between numerical and analytical results in (c) is
  $\lesssim 5\%$ and decreases with the decrease of $(N-N_c)/N_c>0$.
In simulations the initial conditions in the spatial Gaussian form as described in Section \ref{section:simulationalgorithm}.
  }
\label{fig:Lthirdorder}
\end{figure}

Our {\it second main result} is that the asymptotic solution
of~\eqref{ataufullnextorder} in the limit $t\to t_c$, together
with~\eqref{adef1} and~\eqref{taudef1}, is given by
\begin{eqnarray} \label{scaling5}
&L(t)=2e^{-\frac{2+\gamma}{2}}\sqrt{t_c-t}\exp{\Big \{-\sqrt{-\frac{\ln{\beta (t_c-t)}}{2}}+\frac{-1 + b \ln{x}}{2x} }
\nonumber \\
&
\qquad\qquad+\frac{-1 + 2 b+2\tilde M(1-b\ln{x})}{4 x^2}
%
%
+O\left (\frac{1}{x^2}\right )+O\left (\frac{(\ln{x})^2}{x^3}\right ) \Big \}, \nonumber \\
  &x=\sqrt{-2\ln{\beta (t_c-t)}}-\tilde M, \nonumber \\
&\tilde M=  -2 - \gamma + \ln{2}, \nonumber \\
  &b=1+\frac{\pi^2}{6} ,   \\
&\beta =2\exp{\left \{2 l^* -\frac{\tilde M^2}{2}  \right \}}, \nonumber \\
 &l^*=-\ln{L_0} -\frac{1}{4}\ln^2{a_0}+\frac{\tilde M+1}{2}\ln{a_0}-\frac{b}{2}\left ( \ln{\ln{\frac{1}{a_0}}}+\frac{1}{\ln{\frac{1}{a_0}}} \right ), \nonumber \\
 &L_0: =L(t_0), \quad a_0:=a(t_0)=-L(t_0)\partial_t L(t_0). \nonumber
\end{eqnarray}
This scaling was presented without derivation in
Ref.~\cite{DyachenkoLushnikovVladimirovaAIP2011}.  The time
$t=t_0<t_c$ is chosen arbitrarily, provided that at $t=t_0$, the
solution is close to the self-similar
form~\eqref{pselfsimilarfull}. (More details about choice of $t_0$ are given at the end of
 Section \ref{section:amplitudeexpansion} and in Figure \ref{fig:Lthirdorder}.)  It is seen from~\eqref{scaling5} that $L(t)$ depends
on the initial values $L(t_0)$ and $\partial_tL(t_0)$.  The order of
error terms in~\eqref{scaling5} are discussed below, after
Eq.~\eqref{LtY0}.

\begin{figure}
\begin{center}
\includegraphics[width = 2.98 in]{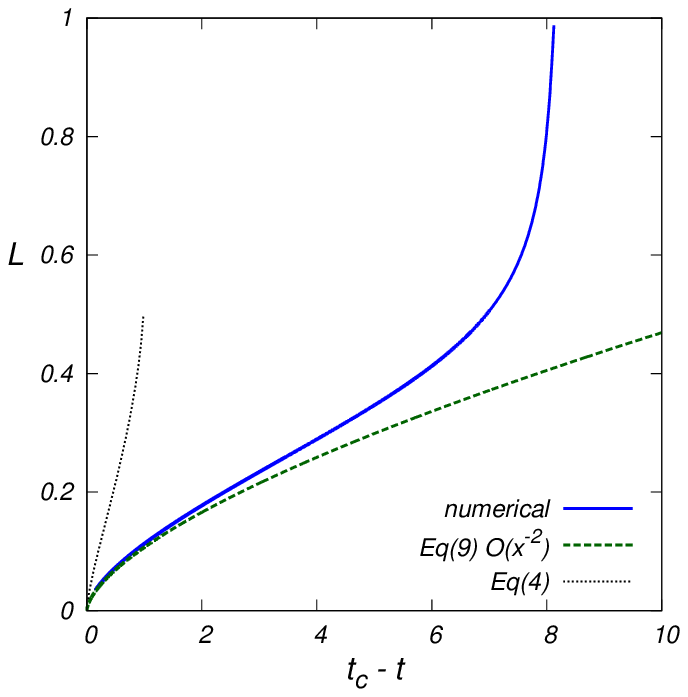}
\end{center}
\caption{Dependence $L(t)$ during the time interval significantly
  exceeding the time interval of self-similar regime.  The solid line
  shows the results of the numerical simulations for $N= 1.0250N_c$,
  the dotted line shows the scaling~\eqref{Ltexplog2}, while the
  dashed line shows the scaling~\eqref{scaling5} with all terms up to
  $O(x^{-2})$.  The six-fold decrease of $L$ from the initial value
  $L(0)=0.98773$ already gives a good agreement between numerical simulation
  and~\eqref{scaling5}, with relative difference between them
  $\lesssim 7\%$ for $L<0.15$.  The scaling~\eqref{Ltexplog2} agrees
  with simulation only in order of magnitude for $L\simeq 0.15$.
  Figure~\ref{fig:Lthirdorder}c
  shows the same curves for $N= 1.0250N_c$ zoomed-in to the origin.
  }
\label{fig:Lthirdorderfull}
\end{figure}

Our {\it third main result} is the comparison of~\eqref{scaling5} with
direct numerical simulations of RKSE.
Figure~\ref{fig:Lthirdorder} shows excellent agreement between the
theory and simulations.  In the limit $t \to t_c$, the new
scaling~\eqref{scaling5} reduces to~\eqref{Ltexplog2}.  We demonstrate,
however, that while~\eqref{Ltexplog2} is asymptotically correct, it
is in quantitative agreement with both~\eqref{scaling5} and
simulations only for unrealistically small values
\begin{equation}
\label{doubleexponent}
L\lesssim 10^{-10000}.
\end{equation}
In contrast, the scaling~\eqref{scaling5} is accurate starting from a
moderate decrease of $L(t)$ from the initial value $L(0)$.
Figure~\ref{fig:Lthirdorderfull} shows the simulation with $N= 1.0250N_c$, where~\eqref{scaling5} is accurate (with the relative error $\lesssim 7\%$)
for $L(t)/L(0)\lesssim 0.15$.

\subsection{Outline of the paper}

The paper is organized as follows.  In
Section~\ref{section:collapse2D} we consider general properties of
collapses in RKSE and their analogies with the collapses in NLSE.  In Section~\ref{section:selfsimilarcollapse2D} we study a
collapsing self-similar solution of RKSE.  We write the
self-similar solution as a rescaled steady state solution in new
``blow up'' variables.  In these variables, the self-similar solution
transforms into the approximate steady-state solution.  The full
collapsing solution evolves slowly about the steady-state solution,
and depends on the small adiabatically slow parameter $a$ defined
in~\eqref{adef1}.  We use
a gauge transformation to a new dependent
variable for perturbations about the self-similar solution.  The gauge
transformation brings the linearization operator about the
self-similar solution to a self-adjoint form.  In
Section~\ref{section:spectrum} we discuss the spectrum and
eigenfunctions of the linerization operator.  In
Section~\ref{section:amplitudeexpansion} we expand the perturbations
about the self-similar solution into eigenfunctions of the
self-adjoint linearization operator, in order to derive a set of
amplitude equations for the coefficients of the expansion.
Compatibility conditions to ensure
an adiabatic form of the expansion
result in the ODE~\eqref{ataufullnextorder}.
In Section~\ref{section:blowuprate}, we solve
Eq.~\eqref{ataufullnextorder} to derive the scaling~\eqref{scaling5}.
In Section~\ref{section:simulationalgorithm}, we describe the
simulation algorithm and the procedure for the extraction of the
parameters of collapsing solutions from simulations.  In Section
\ref{section:conclusion}, the main results of the paper and future
directions are discussed. In~\ref{appendix:KellerSegelModel}, we provide an extensive
description of the Keller-Segel model and derive the reduced
Keller-Segel equation.  In~\ref{appendix:MeijerGfunction}, we
provide the explicit expressions for the calculation of the scalar
products from Section~\ref{section:amplitudeexpansion}; these
expressions are obtained by using the asymptotic expansions of the
Meijer G-function and the $\Gamma$-function.

\section{Collapse of RKSE and NLSE}
\label{section:collapse2D}

Equation~\eqref{pottscontinuousKellerSegelreduced} has a form of a conservation law
\begin{eqnarray}
\partial _t \rho&=&-\nabla\cdot \Gamma, \label{KellerSegelGamma}
\end{eqnarray}
where $\Gamma$ is the flux of the bacterial  density
given by
\begin{eqnarray}
 \Gamma=-\rho\nabla \Big [\ln \rho-c \Big ], \label{Gammadef1}
\end{eqnarray}
and $c({\bf r})$ is determined by the fundamental solution
$E({\bf  r},{\bf r}')$ of the Poisson equation. The 2D case considered here
implies that
\begin{eqnarray}\label{cdef1poisson}
c({\bf r})= -\int E({\bf r},{\bf r}') \rho({\bf r}')d{\bf r}',  \quad
E({\bf r},{\bf r}')=\frac{1}{2\pi}\ln |{\bf r}-{\bf r}'|.
\end{eqnarray}
Eq.~\eqref{cdef1poisson} allows to rewrite
Eq.~\eqref{pottscontinuousKellerSegelreduced} as a closed
integro-differential equation for $\rho$. The integral term in that equation originates from \eqref{cdef1poisson}  and represents the nonlocality of interaction
due to diffusion of chemoattractant.

Assuming decaying boundary conditions at infinity, we obtain the
conservation of the total number $N$ of bacteria:
\begin{eqnarray}\label{Ndef1section5p1}
N=\int \rho({\bf r}) d {\bf r}=const.
\end{eqnarray}
One can also define a Lyapunov functional
\begin{eqnarray}\label{Edef1}
{\mathcal E}=\int \Big [\rho({\bf r}) \ln \rho({\bf r})-\rho({\bf r}) -\frac{\rho({\bf r}) c({\bf r})}{2}\Big ]  d {\bf r},
\end{eqnarray}
and represent Eq.~\eqref{pottscontinuousKellerSegelreduced} in a gradient form
\begin{eqnarray}\label{gradient1}
\partial_t \rho=\nabla \cdot \left (\rho\nabla \frac{\delta {\mathcal E}}{\delta \rho} \right ), \quad  \frac{\delta  {\mathcal E}}{\delta \rho}=\ln \rho-c,
\end{eqnarray}
where the Lyapunov functional~${\mathcal E}$ is a non-increasing function of time
\begin{eqnarray}\label{Et1}
 \frac{d {\mathcal E}}{d t}=-\int \frac{\Gamma^2}{\rho} d {\bf r}.
\end{eqnarray}
Functional ${\mathcal E}$ is conserved only for a steady state solutions with
zero flux $\Gamma=0.$

Although Eq.~\eqref{pottscontinuousKellerSegelreduced} is
a gradient non-Hamiltonian system (as follows from \eqref{gradient1}),
it has many striking similarities with NLSE \eqref{nls1kellersegel}
which can be written in a Hamiltonian form $i\partial_t
\psi=\frac{\delta H}{\delta \psi^*}$ with the Hamiltonian
\begin{equation}\label{Hdef}
H=\int
\left[ |\nabla\psi|^2-\frac{|\psi|^4}{2}\right ]d{\bf r}.
\end{equation}

To prove existence of collapse in RKSE one can use a
positive-definite quantity $A= \int r^2 \rho d{\bf r}$, which
determines the mean square width of bacterial density
distribution~\cite{BilerWoyczinski1998,ChavanisSirePRE2006}.
Vanishing of $A$ guarantees the existence of collapse because of
conservation of $N$.  The proof of collapse existence for NLSE in $D=2$ is
based on a virial identity ~\cite{VlasovPetrishchevTalanovRdiofiz1971,ZakharovJETP1972}:
\begin{equation}\label{BvirialNLS}
\partial^2_t
B=8H.
\end{equation}
 Here $B:= \int r^2 |\psi|^2 d{\bf r}$ and $H$ is defined in \eqref{Hdef}. If $H<0$ then the positive-definite $B$ turns negative in a finite time as follows from \eqref{BvirialNLS}.
It means that the negative value of the Hamiltonian is
the sufficient condition for the collapse in NLSE.    We also recall that  in
the quantum mechanical interpretation of NLSE, $|\psi|^2$ is the
probability density of number of particles, i.e., the analog of $\rho$ in
RKSE.  Thus $A$ from RKSE is the analog of $B$ in NLSE. However, RKSE is the non-Hamiltonian system and the
direct analogy with a virial theorem for $B$ does not work. Instead
one can calculate a time derivative of $A$ using
Eqs.~\eqref{pottscontinuousKellerSegelreduced}
and~\eqref{cdef1poisson}, integration by parts, and vanishing boundary
conditions at infinity.  For $D=2$, this procedure gives:
\begin{equation}\label{Atkellersegel}
\partial_t A =4N-\frac{1}{2\pi}\int 2{\bf r}\cdot ({\bf r}-{\bf r}')\frac{\rho({\bf r})\rho({\bf r}')}{ |{\bf r}-{\bf r}'|^2}d{\bf r}d{\bf r}'=4N-\frac{N^2}{2\pi}.
\end{equation}
Here we also used symmetrization over ${\bf r}$ and ${\bf r}'$. One
concludes from~\eqref{Atkellersegel} that $A_t<0$ for $N>8\pi$ and
$A$ turns negative in a finite time. That condition defines the critical number of bacteria
\begin{equation}\label{Nccritical}
N_c=8\pi
\end{equation}
because $A<0$ proves  the existence of collapse by contradiction ($A$ is the positive-definite).

The existence of the critical number of bacteria \eqref{Nccritical} is
another similarity with NLSE, where the critical number of
particles $P_{c}=\int |\psi|^2 d{\bf r}\simeq 11.70\ldots$
The difference between collapses in NLSE and RKSE is that according to \eqref{Atkellersegel} for
RKSE, any initial condition with $N>N_c$ develops into the
collapsing solution in a finite time, while for NLSE, $P>P_c$ is
the necessary condition for collapse but not the sufficient condition.
Another qualitative
difference between RKSE and NLSE is that RKSE is the
integro-differential equation while NLSE is a
partial differential equation (PDE).  However, it was shown in
Refs.~\cite{LushnikovPRA2002,LushnikovPRA2010} that the generalized
virial identity allows to prove the collapse existence in an
integro-differential equation of NLSE-type with nonlocal nonlinearity.
This type of nonlinearity describes, e.g., Bose-Einstein condensate with nonlocal
dipole-dipole interaction.  Collapse of such condensate was recently
achieved in experiment~\cite{LahayeMetzPfauEtAlPRL2008}.

Qualitative similarities between collapses in RKSE and in NLSE can
be also understood if we recall that RKSE is the mean-field
approximation for the dynamics of self-gravitating gas of brownian
particles, while NLSE is the mean-field approximation for the quantum
dynamics of atoms with Bose statistics and attraction at ultra-cold
temperatures.
Thus both RKSE and NLSE approximate the dynamics of gas of particles
with attraction. The principle difference is that the dynamics of
brownian particles (RKSE) is diffusive (originates the overdamped
motion with random force), while the dynamics of Bose atoms is the quantum
analog of Newtonian mechanics. In both cases collapse occurs if the
number of particles is large enough to cause attraction overcoming
either quantum pressure (NLSE) or diffusion (RKSE).

\section{Self-similar collapsing solution of the 2D reduced Keller-Segel equation}
\label{section:selfsimilarcollapse2D}

2D RKSE~\eqref{pottscontinuousKellerSegelreduced}
is invariant under the scaling transformations $\rho({\bf
  r},t)\to \frac{1}{L^{2}}\rho(\frac{1}{L}{\bf r} ,\frac{1}{L^{2}}t)$,
$c({\bf r},t)\to c(\frac{1}{L^{}}{\bf r} ,\frac{1}{L^{2}}t)$
for any $L(t)\equiv L=const>0$.
Similar property holds for
NLSE.  The 2D RKSE has a static, radially-symmetric solution
\begin{equation}\label{pstaticsection5p1}
\begin{split}
& \rho_0=\frac{8}{(1+r^2)^2},  \\
 & c_0=-2 \ln (1+r^2),
 \end{split}
\end{equation}
which corresponds to the critical number of bacteria, $N(\rho_0)=N_c=8\pi$.
This property is another striking similarity with the ground state
soliton solution $\psi=R(r)e^{it}, \ R(r)\ge 0$ of NLSE
containing exactly the critical number of
particles, $P_{c}=\int R^2 d{\bf r}.$

Assume that collapse is centered at $r=0$. Then the solution of RKSE in the limit $t\to t_c$ approaches a radially-symmetric,
self-similar solution.  The self-similar solution has the
form of the rescaled stationary solution~\eqref{pstaticsection5p1}
 with a time-dependent scale (the collapse width) $L(t)$:
\begin{equation}\label{pselfsimilar}
\begin{split}
& \rho(r,t)=\frac{1}{L(t)^{2}}\rho_0\left ( \frac{r}{L(t)} \right),   \\
 & c(r,t)=c_0\left (  \frac{r}{L(t)}\right ).
  \end{split}
 \end{equation}
The scale $L(t)$ approaches zero for $t \to t_c$.

To describe the radially-symmetric solution
we introduce the new dependent variable $m$ as follows,
   \begin{equation}\label{mdef}
   m(r,t)=\frac{1}{2\pi}\int_{|r'|\le r}\rho({\bf r}',t)\ d{\bf r}',
   \end{equation}
which allows us to rewrite RKSE as the
closed equation for $m$~\cite{HerreroVelazquezMathAnn1996}:
   \begin{equation}\label{meq1}
\partial_t m=r\partial_r r^{-1}\partial_r m+r^{-1} m\partial_r m.
 \end{equation}
Here, $m(r,t)$ has the meaning of the mass (the number of bacteria)
inside the circle of radius $r$ (up to a factor $2\pi$). Boundary
condition for $m$ at $r\to \infty$ is simply related to the total
number of bacteria: $m|_{r=\infty}= N/(2\pi)$.  In contrast to RKSE,
Eq.~\eqref{meq1} is PDE for $m$. This simplification is possible only for
radially-symmetric solutions of RKSE.

In terms of $m$, the steady state
solution~\eqref{pstaticsection5p1} of RKSE takes the
following form:
\begin{equation}\label{pstaticm0}
m_0=\frac{4r^2}{1+r^2},
\end{equation}
and the self-similar solution~\eqref{pselfsimilar} becomes
\begin{eqnarray}\label{mselfsimilar}
m_{selfsimilar}=\frac{4y^2}{1+y^2}, \nonumber \\
y=\frac{r}{L}.
\end{eqnarray}
The boundary condition at infinity gives the critical number of bacteria,
$2\pi m_{selfsimilar}\big |_{y\to \infty}\to 8\pi=const$, It
also indicates that bacterial collapse is strong because the
number of bacteria trapped within the collapse is nearly constant.

Assuming a power law dependence $L(t)\propto (t_0-t)^\beta$ of the
collapse width in the
self-similar solution~\eqref{mselfsimilar} one concludes that all
terms in Eq.~\eqref{meq1} are of the same order provided $\beta=1/2$,
which is similar to NLSE where also the collapsing width  $\propto (t_0-t)^{1/2}$.
Like for NLSE, the self-similar solution~\eqref{mselfsimilar} is not an exact
solution of Eq.~\eqref{meq1}.  To account for the difference, it
is necessary to consider the logarithmic correction to $L(t)\sim
(t_0-t)^{1/2}$: $L= (t_0-t)^{1/2}f(\ln{(t_0-t)})$, where
$f(\ln{(t_0-t)})$ is a slow function compared with $(t_0-t)^{1/2}$.
This slow function comes from the nearly exact balance between linear and
nonlinear terms of RKSE (between diffusion and attraction).
The same  slow function allows
to introduce a small parameter $a$, defined in~\eqref{adef1},
which is a slow function of $(t_0-t)^{1/2}$ compared with $L$.  The
balance between linear and nonlinear terms of RKSE improves with
decrease of $a\to 0$.

Based on the analogy with the critical NLSE, we introduce in
Eq.~\eqref{meq1} the new independent ``blow up'' variables~\cite{DLS2009}:
 \begin{equation}\label{taudef1a}
 \begin{split}
& y=\frac{r}{L},  \\
& \tau=\int^t_0 \frac{dt'}{L(t')^2}.
\end{split}
\end{equation}
These new variables transform Eq.~\eqref{meq1} into
 the equation for a new unknown function
\begin{equation}\label{varphidef}
\varphi(y,\tau)\equiv m(r,t)
\end{equation}
into the following equation
\begin{equation}
\partial_\tau
\varphi=y\partial_y (y^{-1}\partial_y\varphi)+y^{-1}\varphi\partial_y\varphi-a
y\partial_y\varphi, \label{eqn:NLHBlowup}
\end{equation}
where $a$ is given by~\eqref{adef1}. 
The advantage of working in blow up variables is that the
collapse occurs at $\tau=\infty$ instead of
$t = t_c$, so that the collapse time
$t_c$ is eliminated from consideration.
Also, the function $\varphi$ has bounded derivatives.

\begin{figure}
\begin{center}
\includegraphics[width = 5.98 in]{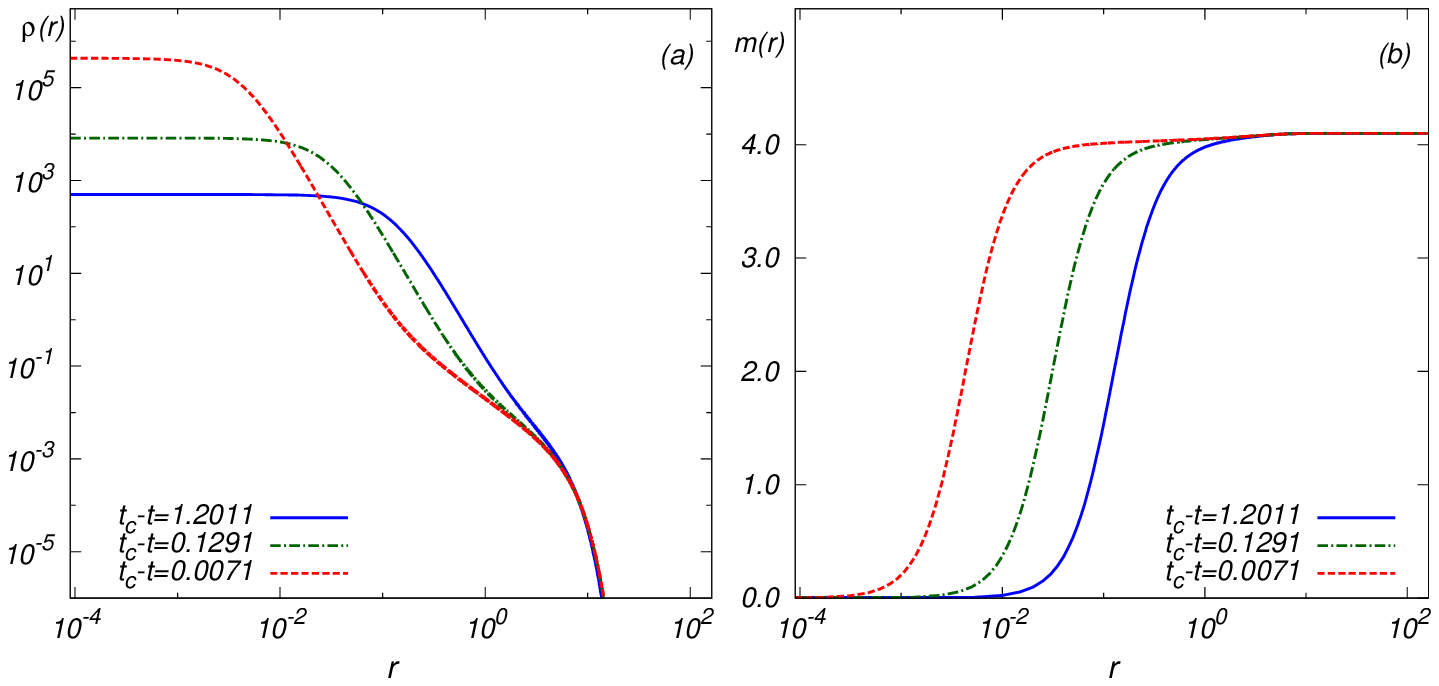}\\
\includegraphics[width = 5.98 in]{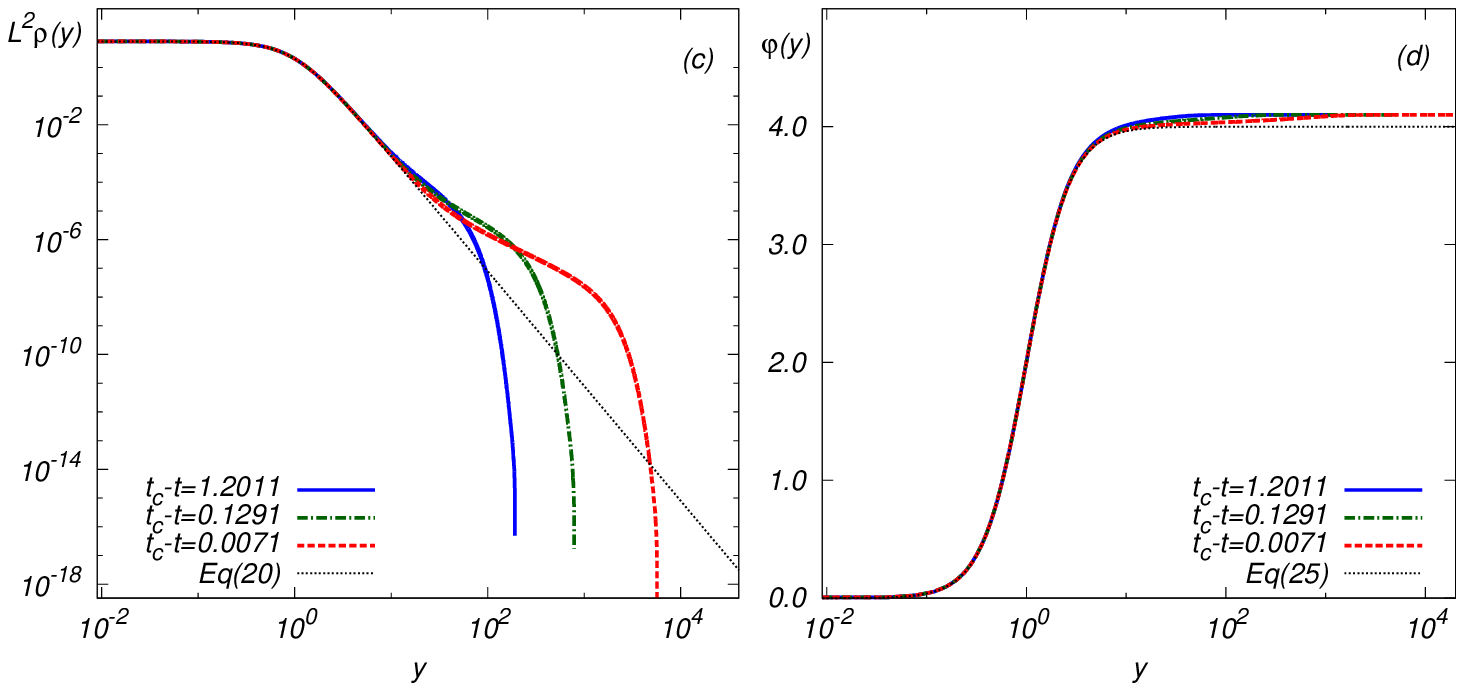}
\end{center}
\caption{ The spatial dependence of the density $\rho$ (panels a,c) and the mass $m$
  (panels b,d) at different moments of time for the simulation with
  $N=1.0250N_c$.  In the top row (panels a,b), the data is shown in
  simulation coordinates.  In the bottom row (panels c,d), rescaled
  density, $L^2 \rho(y)$, and mass, $\varphi(y,\tau)\equiv m(r,t)$,
  are shown as functions of rescaled radius, $y=r/L$.  In panel (a),
  notice the growth of $\rho(r)$ near the origin and a nearly steady tail.  In
  panel (c), notice the convergence to the static solution
  \eqref{pstaticsection5p1},\eqref{pselfsimilar} in the growing
  neighborhood of $y=0$.  In loglog scale the deviation from that
  static solution has the form of a bump.  The bump moves away from the
  origin as $t \to t_c$.
}
\label{fig:rhor}
\end{figure}

Figures~\ref{fig:rhor}a,b shows that as $t\to t_c$, the density  $\rho(r)$  grows near
$r=0$ while the tail of $\rho(r)$ is practically frozen for
$r\gtrsim 3$ (on a timescale of collapse).
In contrast, the solution in the blow up variables is steady at
$y\lesssim 1$ and is well-approximated
by~\eqref{pstaticsection5p1} and~\eqref{pselfsimilar}, as shown in
Figure~\ref{fig:rhor}c,d. It is also seen that the deviation of solution from ~\eqref{pstaticsection5p1} moves away from the origin
  $y=0$ as $t\to t_c$.

Based on our assumption that $a$ is a slow function, it is natural to
look at the solutions of Eq.~\eqref{eqn:NLHBlowup} in the adiabatic
approximation where one can neglect $\tau$-derivative in the left-hand
side (lhs) of Eq.~\ref{eqn:NLHBlowup}. Then, assuming that $|a|\ll 1$, one
can expand the solution of~\eqref{eqn:NLHBlowup} in powers of $a$
starting from~\eqref{mselfsimilar} for the power zero.  Unfortunately,
the term $-a y\partial_y\varphi$ grows with $y$ and violates the
expansion for large $y$.  So, the adiabatic approximation can only
work locally and is restricted to not very large $y$, a situation which is
familiar from the analysis of collapse in NLSE.   This however does not create a problem
because the behavior at large $y$ does not affect the self-similar
solution near zero.

It is convenient to present a general solution of
Eq.~\eqref{eqn:NLHBlowup} in the following form~\cite{DLS2009}:
 \begin{equation}\label{vdef1}
\varphi(y,\tau)=\frac{4y^2}{1+y^2}+e^{\frac{a}{4}y^2}\frac{y^2}{y^2+1}v(y,\tau),
\end{equation}
where $v(y,\tau)$ includes
all corrections with respect to
the self-similar solution~\eqref{mselfsimilar}.  Here, the factor
$e^{\frac{a}{4}y^2}$ (which plays a role of a gauge transform) is
inspired by a somewhat similar factor $e^{-i\frac{a}{4}y^2}$ in
the self-similar solution of NLSE~\cite{SulemSulem1999}.
However, the absence of $-i$ in the exponent makes RKSE case
quite distinct from NLSE case.

Substitution of~\eqref{vdef1} into~\eqref{eqn:NLHBlowup} gives the
following equation:
\begin{equation} \label{veq1}
\partial_\tau v+\hat {\mathcal{L}}_av=F.
\end{equation}
Here
\begin{equation}\label{Ldef01}
\hat{\mathcal{L}}_a=-\frac{1}{y^3}\partial_y y^3\partial_y
-\frac{8}{(1+y^2)^2}+\big [\frac{a^2}{4}y^2-2a+\frac{2a}{1+y^2}
\big ]
\end{equation}
is the linear operator corresponding to the linearization
of~\eqref{eqn:NLHBlowup} with respect to~\eqref{mselfsimilar}. The
right-hand side,
\begin{equation} \label{Fdef1section5p1}
F=-\frac{\partial_\tau a}{4}y^2 v-\frac{8a}{y^2+1}e^{-a y^2/4}+\frac{ay^2
v^2 }{2(y^2+1)}e^{a y^2/4}+\frac{2 v^2 }{(y^2+1)^2}e^{a
y^2/4}+\frac{yv \partial_yv }{y^2+1}e^{a y^2/4},
\end{equation}
is responsible for all other terms. These other terms include terms nonlinear in $v$,
inhomogeneous terms, and linear terms.
Generally, $F$ cannot be zero because~\eqref{mselfsimilar} is not an
exact solution of~\eqref{eqn:NLHBlowup} for nonzero $a$.  Notice that
up to now we have not made any approximations, so
Eqs.~\eqref{vdef1}--\eqref{Fdef1section5p1} are equivalent to
Eq.~\eqref{eqn:NLHBlowup}.

The advantage of the definition~\eqref{vdef1} is that the operator
${\hat {\mathcal{L}}}_a=-\frac{1}{y^3}\partial_y y^3\partial_y+V(y)$ has the
form of the radially symmetric Schr\"odinger operator in spatial
dimension four ($D=4$) with the potential
\begin{equation} \label{Vpotentialdef}
 V(y)=-\frac{8}{(1+y^2)^2}+\big[ \frac{a^2}{4}y^2-2a+\frac{2a}{1+y^2}
\big].
\end{equation}
It means that  ${\hat {\mathcal{L}}}_a$  is the self-adjoint operator with the scalar product
\begin{equation} \label{scalarproduct1}
\langle \psi ,\phi \rangle =\int\limits^{\infty}_{0} \psi(y) \phi(y) \,  y^3 \, dy.
 \end{equation}
The potential $V(y)\to \infty $ for $y\to \infty$, which ensures that
${\hat {\mathcal{L}}}_a$ has only discrete spectrum.  This allows us to expand
arbitrary $v$ in a discrete set of eigenfunctions of $\hat
{\mathcal{L}}_a$:
\begin{equation} \label{vexp1}
v=c_1\psi_1+c_2\psi_2+c_3\psi_3+\ldots,
\end{equation}
where $c_1(\tau), \ c_2(\tau), \ \ldots$ are $\tau-$dependent
coefficients of the expansion (below we often omit argument $\tau$ for
brevity), $\psi_j(y)$ are the eigenfunctions of $\hat
{\mathcal{L}}_a$,
\begin{equation} \label{eigenvalfunc01}
\hat {\mathcal{L}}_a\psi_j=\lambda_j \psi_j,
\end{equation}
and $\lambda_j$ are the respective eigenvalues.  The eigenvalues are
ordered starting from the lowest eigenvalue as
$\lambda_1<\lambda_2<\ldots$. All eigenvalues are real and
non-degenerate as discussed in the next section.

Note that the use of the scalar product~\eqref{scalarproduct1} (which
corresponds to the radially symmetric Schr\"odinger operator in $D=4$)
is simply an auxiliary mathematical trick,  which is effective
because the operator $\hat {\mathcal{L}}_a$ is self-adjoint with this
scalar product.  We remind that all solutions obtained below
correspond to RKSE~\eqref{pottscontinuousKellerSegelreduced} with
$D=2$.

\section{Spectrum of linearization operator}
\label{section:spectrum}

The eigenvalues~\eqref{eigenvalfunc01} of the linearization operator
$\hat{\mathcal{L}}_a$ are given by the following implicit expression,
\begin{equation}
\frac{\lambda+2a}{2a}\lsb \ln\frac{1}{a}-\Psi\lb -\frac{\lambda}{2a} \rb+K \rsb=1+\O{a^{1/2}\ln\frac{1}{a}},
\label{eqn:eigenvalue}
\end{equation}
as it was proven in
Ref.~\cite{DejakLushnikovOvchinnikovSigalPhysD2012} using a rigorous
version of the method of matched asymptotics.  Here $K:=\ln
2-1-2\gamma$, while $\Psi$ is the digamma function, defined as
$\Psi(s)=\frac{d}{ds}\ln \Gamma(s)$, where $\Gamma(s)$ is the gamma
function.

Solving~\eqref{eqn:eigenvalue} for $\lambda$ gives the spectrum of
$\hat {\mathcal{L}}_a$, starting from the lowest eigenvalues, as follows
\begin{equation}\label{lambtot}
\begin{split}
&\lambda_1=a\left (-2+\frac{2}{\ln{\frac{1}{a}}}+2(1+\gamma-\ln{
2})\frac{1}{(\ln{\frac{1}{a}})^2}+\left [2(K+\gamma)^2-\frac{\pi^2}{3}\right ]\frac{1}{(\ln{\frac{1}{a}})^3}\right ) \\
&\qquad\qquad\qquad\qquad\qquad\qquad\qquad\qquad\qquad\qquad\qquad\qquad+O\left (\frac{a}{(\ln{\frac{1}{a}})^4}\right ),
 \\
&\lambda_2=a\left (\frac{2}{\ln{\frac{1}{a}}}+2(2+\gamma-\ln{
2})\frac{1}{(\ln{\frac{1}{a}})^2}+\left[-4+2(K+\gamma)^2-\frac{\pi^2}{3}\right ]\frac{1}{(\ln{\frac{1}{a}})^3}\right )\\
&\qquad\qquad\qquad\qquad\qquad\qquad\qquad\qquad\qquad\qquad\qquad\qquad+O\left (\frac{a}{(\ln{\frac{1}{a}})^4}\right ), \\
&\lambda_3=a\left (2+\frac{2}{\ln{\frac{1}{a}}}+(5+2\gamma-2\ln{
2})\frac{1}{(\ln{\frac{1}{a}})^2}\right . \\
&\qquad\qquad\left .+\left[2(1-3K-3\gamma)+2(K+\gamma)^2-\frac{\pi^2}{3}\right ]\frac{1}{(\ln{\frac{1}{a}})^3}\right )
+O\left (\frac{a}{(\ln{\frac{1}{a}})^4}\right ), \\
&\ldots 
\end{split}
\end{equation}
Eigenfunctions $\psi_j$ can be also approximated from the method of
matched asymptotics.

In Section~\ref{section:amplitudeexpansion} we need to calculate multiple
integrals which involve $\psi_j$.  For this purpose, it is more
convenient to use the variational approximation  for eigenfunctions
obtained in Ref.~\cite{LushnikovPhysLettA2010}:
\begin{equation}\label{eigvector01tot}
\begin{split}
&\tilde \psi_1=\frac{8}{1+y^2}e^{-ay^2/4}, \\
&\tilde \psi_2=\frac{8}{1+y^2}\left(1+
\frac{a y^2}{2}- \frac{ ay^2}{2}\ln{(1+y^2)}\right )e^{-ay^2/4},  \\
&\tilde\psi_3=\frac{8}{1+y^2}\left(1+
a y^2\left [-\frac{\pi^2 \ln{\frac{1}{a}}}{\pi^2-12}+ \frac{-12+\pi^2(2+\gamma-\ln{2})}{\pi^2-12}\right ]  \right . \\
&    \left . \quad \quad +ay^2\ln{(1+y^2)}\frac{12}{\pi^2-12} +\frac{a^2y^4}{4} \left [\ln{\frac{1}{a}}-3-\gamma+\ln{2}-\frac{24}{\pi^2-12} \right ] \right ) e^{-ay^2/4},
\end{split}
\end{equation}
where $\tilde \psi_j$ means the variational approximation to $\psi_j$,
$j=1,2,\ldots$.  We estimate the accuracy of the variational
approximation by calculating the variational approximation for the
three lowest eigenvalues $\lambda_1$,  $\lambda_2$, and $\lambda_3$ as
$\lambda_j=\frac{\langle\tilde \psi_j,\hat{\mathcal{L}}_a\tilde\psi_j
  \rangle}{\langle\tilde \psi_j,\tilde\psi_j \rangle}, \ j=1,2,3.$
These scalar products involve the calculation of integrals of the type
described in~\ref{appendix:MeijerGfunction}.  Expansion of the
resulting expressions for integrals $\lambda_1$,  $\lambda_2$, and
$\lambda_3$ in inverse powers of $\ln{\frac{1}{a}}$ agrees with exact
results~\eqref{lambtot} up to order $O\left
(\frac{1}{(\ln{\frac{1}{a}})^2} \right )$ for $\lambda_1, \ \lambda_2$
and up to order $O\left (\frac{1}{\ln{\frac{1}{a}}} \right )$ for
$\lambda_3$. This is the best result we are able to achieve with the
variational approximation.  This accuracy will be however  sufficient to
obtain~\eqref{ataufullnextorder}.

\section{Amplitude equations}
\label{section:amplitudeexpansion}

Similar to~\eqref{vexp1}, we expand $v$ from~\eqref{veq1} in a set of
approximate variational
eigenfunctions $\tilde \psi_j$, $j=1,2,\ldots$ as follows,
\begin{equation} \label{vexp1var}
v=\sum\limits_{j=1}^\infty c_j\tilde \psi_j,
\end{equation}
where $c_j(\tau)$ are the coefficients of the expansion.  In this Section
we derive a set of amplitude equations for $c_1(\tau), \ c_2(\tau),
\ldots$ from~(\ref{vexp1var}) which provide a solution of  Eq. (\ref{veq1}).
We solve
the amplitude equations exploiting the fact that, at the leading
order in $a$, the solution of (\ref{eqn:NLHBlowup}) is given by
(\ref{mselfsimilar}). (We used that fact in the definition of (\ref{vdef1})).
 We expand all expressions below in integer powers of the small parameters
$a$ and $\frac{1}{\ln{\frac{1}{a}}}$, keeping the lowest nontrivial order
of $a$ and several orders of $\frac{1}{\ln{\frac{1}{a}}}$.

We assume the approximate orthogonality of the variational functions,
\begin{equation} \label{psiorthog}
\langle \tilde\psi_i,\tilde\psi_j\rangle=O(a)\|\tilde\psi_i\|\|\tilde\psi_j\| \quad \mbox{for} \quad i\neq j,
\end{equation}
where $\|\tilde\psi_i\|:=\langle
\tilde\psi_i,\tilde\psi_i\rangle^{1/2}, \ i=1,2,\ldots$. Then, the scalar
multiplication of~\eqref{veq1} onto $\tilde \psi_j$ (with the scalar
product \eqref{scalarproduct1}) results in
\begin{equation}\label{psiLpsi}
\begin{split}
&\langle\tilde\psi_j,\partial_\tau v\rangle+\langle\tilde \psi_j, \hat {\mathcal{L}}_av\rangle-\langle\tilde\psi_j,F(v)\rangle\\
&=\|\tilde\psi_{i} \|^{2}\partial_\tau c_j+\sum_{i=1}^\infty c_i\langle\tilde\psi_j,\partial_\tau \tilde\psi_i\rangle +\sum_{i=1}^\infty c_i\langle\tilde\psi_j,\hat{\mathcal L}_a\tilde\psi_i\rangle-\langle\tilde\psi_j,F(v)\rangle=0. \\
\end{split}
\end{equation}
Here, we neglect corrections from nonexact
orthogonality~\eqref{psiorthog} because, as we show later, these
corrections are of the next order in $a$ when compared with other
terms in~\eqref{psiLpsi}.  In this section, all calculations of scalar
products for~\eqref{psiLpsi} are based on integrals defined
in~\ref{appendix:MeijerGfunction} for the variational
functions~\eqref{eigvector01tot}.  For instance, the direct
calculation for the variational
functions~\eqref{eigvector01tot}  gives the following expressions:
\begin{equation}\label{psinorms}
\begin{split}
&\|\tilde\psi_1\|^2=-32 \ln{a} +32(-1- \gamma+\ln{2})+O\left(a\ln{a}\right ),\\
&\|\tilde\psi_2\|^2=32(\ln{a})^2+32(1+2 \gamma-2\ln{2})\ln{a},\\
&+\frac{16}{3} \left[\pi ^2+6 (\ln{2}-1) \ln{2}+6 \gamma  (\gamma +1-2\ln{2})\right]+O\left(a\ln{a}\right),\\
&\|\tilde\psi_3\|^2=64(\ln{a})^2+\frac{32 \left(-108-48 \gamma+13 \pi ^2+4 \gamma \pi ^2+48 \ln{2}-4 \pi ^2 \ln{2}\right)}{\left(-12+\pi ^2\right) }\ln{a}\\
&+\frac{1}{\left(-12+\pi ^2\right)^2}32 \Big [2 \gamma^2 \left(-12+\pi ^2\right)^2+\pi ^4 [23+\ln{2} (-13+2\ln{2})] )\\
&-24 \pi ^2 [13+\ln{2} (-11+2\ln{2})]+144 (-1+\ln{2}) (-7+2\ln{2})\\
&-\gamma \left(-12+\pi ^2\right) [108-48 \ln{2}+\pi ^2 (-13+4\ln{2})]\Big]+O\left(a(\ln{a})^2\right ).
\end{split}
\end{equation}

We assume (based, e.g., on numerical simulations
in~\cite{BrennerConstantinKadanoff1999,BettertonBrennerPRE2001} and
following Ref.~\cite{LushnikovPhysLettA2010}) that $a$ is the
adiabatically slow function of $\tau$: $\partial_\tau a\ll a^2$.  As
mentioned above, we expand all quantities in the small parameters $a$
and $\frac{1}{\ln{\frac{1}{a}}}$ (it is also seen
in~\ref{appendix:MeijerGfunction} that all integrals involved
in~\eqref{psiLpsi} expand into these parameters) keeping only a leading
order in $a$ and many enough terms in powers of
$\frac{1}{\ln{\frac{1}{a}}}$.
Then the adiabatic assumption $\partial_\tau
a\ll a^2$  requires $\partial_\tau a = a^2
O\left(\frac{1}{\ln{\frac{1}{a}}}\right )$.  We introduce a normalized
function $\tilde a_\tau:=\frac{\partial_\tau
  a}{a^2}\frac{1}{\ln{\frac{1}{a}}}=O(1)+O(a)$.
This
allows to write $\partial_\tau a$ as an expansion in inverse powers of
$\ln{\frac{1}{a}}$ only:
\begin{equation} \label{atau}
\partial_\tau a=a^2\, \frac{1}{\ln{\frac{1}{a}}} \ \tilde a_\tau, \quad \tilde a_\tau=\tilde
a_\tau^{(0)}+\tilde a_\tau^{(1)}\frac{1}{\ln{\frac{1}{a}}}+\tilde a_\tau^{(2)}\frac{1}{\left (\ln{\frac{1}{a}}\right )^2}+O\left (\frac{1}{(\ln{\frac{1}{a}})^3}\right ),
\end{equation}
where the coefficients $\tilde a_\tau^{(0)}, \ \tilde a_\tau^{(1)}$
and $\tilde a_\tau^{(2)}$ are $O(1)$ and do not depend on $\tau$ in
the adiabatic approximation.  Note that the subscript $\tau$ in these
coefficients is {\it not} a partial derivative but rather indication
that these are the expansion coefficients for $\tilde a_\tau$.

Assume that the expansion coefficients $c_1, \, c_2, \ c_3,
\ldots $ in (\ref{vexp1var}) are initially  $O(1)$.  A series expansion of
equations \eqref{psiLpsi} over small $a$,
 using Eq. (\ref{atau}) and
dividing each $j$th equation by $\|\tilde\psi_j\|^2$, together with
\eqref{psinorms}, result at the leading order in the following expressions
\begin{equation}\label{psi1scalarall}
\begin{split}
&\partial_\tau c_1+a-2ac_1+O\left (\frac{a}{\ln{\frac{1}{a}}}\right )=0,  \\
&\partial_\tau c_2+O\left (\frac{a}{\ln{\frac{1}{a}}}\right )=0,  \\
&\partial_\tau c_3+2ac_3+O\left (\frac{a}{\ln{\frac{1}{a}}}\right )=0,  \\
&\partial_\tau c_4+4ac_4+O\left (\frac{a}{\ln{\frac{1}{a}}}\right )=0.  \\
&\ldots
\end{split}
\end{equation}
Here, the terms $2a(j-2)c_j,$ $j=1,2,3$ originate from eigenvalues
for $\tilde\psi_j$  (see Eq.~\eqref{lambtot}), while the
term $a$ in the first equation comes from the scalar product of
$\tilde\psi_1$ with the second term in the right-hand side of
(\ref{Fdef1section5p1}).
Also the contribution from $\partial_\tau\psi_j=(\partial_\tau
a)\partial _a \psi_j, \ j=1,2,\ldots$ is included into $O(\ldots)$
term.  It follows from Eqs.~\eqref{psi1scalarall} that the coefficient
$c_3$ initially decays exponentially (because $a>0$) until it reaches
the adiabatic, quasi-steady state with $c_3 =O\left
(\frac{1}{\ln{\frac{1}{a}}}\right)$.  Our conjecture is that the other
coefficients, $c_4,c_5,\ldots$, also decay exponentially (they
correspond to the larger values $\lambda_{j}$, so that they are
assumed to decay as $c_j\propto \exp{[-2a(j-2)\tau]}$, according to
the linear terms in \eqref{veq1}).    The lack of explicit
expressions for $\tilde \psi_j, \ j\ge 4$ does not allow us to prove this
statement.  We conclude that, after an initial transient, the coefficients $c_3,
\, c_4, \ldots$ reach the adiabatic state with their values
\begin{equation} \label{cqeqall}
  c_3, \, c_4, \ldots =O\left (\frac{1}{\ln{\frac{1}{a}}}\right ).
  \end{equation}
Below we assume this adiabatic state.

In the first equation of (\ref{psi1scalarall}) we assume that
\begin{equation} \label{cqeq1}
c_1=\frac{1}{2}+O\left (\frac{1}{\ln{\frac{1}{a}}}\right )
\end{equation}
to avoid exponential growth of $c_1$ in $\tau.$ (Such artificial
exponential growth would result in error in estimating $t_c$.)

We have now a freedom in selecting $c_2$, and we choose it so that $v\to 0$ for
any $y$ as $a\to 0.$ According to (\ref{eigvector01tot}),
$\tilde\psi_1(y)|_{y=0}=\tilde\psi_2(y)|_{y=0}=8$ so we set
\begin{equation} \label{cqeq2}
c_2=-\frac{1}{2}+O\left (\frac{1}{\ln{\frac{1}{a}}}\right ).
\end{equation}
In this case, $c_1\tilde\psi_1+c_2\tilde \psi_2=O(a)$ for $y=O(1)$,
i.e.  $v$ in~\eqref{vdef1}
vanishes with $a\to 0$, as
we expect from the self-similar solution~\eqref{mselfsimilar}.

Equations.~\eqref{cqeqall},~\eqref{cqeq1}, and \eqref{cqeq2} justify the
adiabatic approximation, which means that the coefficients $c_1, \,
c_2, \, c_3, \, c_4, \ldots$ depend on $\tau$ only through $a$, and one
can expand them in series of
inverse powers of $\ln{\frac{1}{a}}$:
\begin{equation} \label{cexp1}
\begin{split}
& c_1=\ \ \frac{1}{2}+\sum \limits_{k=1}^{\infty}d_1^{(k)} \frac{1}{(\ln{\frac{1}{a}})^k}+O(a),
\\
& c_2=-\frac{1}{2}+\sum \limits_{k=1}^{\infty}d_2^{(k)} \frac{1}{(\ln{\frac{1}{a}})^k}+O(a),
\\
& c_3=\quad \quad \quad \sum \limits_{k=1}^{\infty}d_3^{(k)} \frac{1}{(\ln{\frac{1}{a}})^k}+O(a),
\\
& \ldots
\end{split}
\end{equation}
where the expansion coefficients $d_i^{(j)} =O(1)$ for any $i, \,j$;
the coefficients do not explicitly depend on $\tau$ in the adiabatic
approximation.

It follows from (\ref{cexp1}) and (\ref{atau})  that
\begin{eqnarray} \label{ctauall}
\partial_\tau c_j= O \left  (\partial_\tau \frac{1}{\ln{\frac{1}{a}}}\right )=O\left (\frac{a}{(\ln{\frac{1}{a}})^3}\right ), \quad j=1,2,3,\ldots.
\end{eqnarray}

Similar to derivation of Eqs. (\ref{psi1scalarall}), we now perform a series expansion of equations \eqref{psiLpsi}  into small $a$
(but in contrast to
the derivation of Eqs. (\ref{psi1scalarall}) we proceed to the higher
orders of expansion) using
Eqs. \eqref{psinorms},(\ref{atau}),(\ref{cexp1}) to obtain following
equations:
\begin{eqnarray} \label{c1tot1}
&\partial_\tau c_1 + \frac{a}{\ln{\frac{1}{a}}} \left
[\frac{\tilde a_\tau^{(0)}}{2}-2d_1^{(1)}\right
]
\nonumber \\
&
\quad
+\frac{a}{(\ln{\frac{1}{a}})^2} \left
[\frac{\tilde a_\tau^{(1)}}{2}+2d_1^{(1)}-2d_1^{(2)}+2d_2^{(1)}
-\tilde a_\tau^{(0)}d_2^{(1)}+2d_3^{(1)}\right
]\nonumber \\
&\qquad \qquad  \qquad \qquad \qquad\qquad \qquad \qquad \qquad +O\left (\frac{a}{(\ln{\frac{1}{a}})^3}\right )=0,
\\
& \partial_\tau c_2 + \frac{a}{\ln{\frac{1}{a}}} \left
[-1-\frac{\tilde a_\tau^{(0)}}{2}\right
]
\nonumber \\
&
 +\frac{a}{(\ln{\frac{1}{a}})^2} \left
[-1-\frac{\tilde a_\tau^{(1)}}{2}+2d_2^{(1)}
+\tilde a_\tau^{(0)}(d_2^{(1)}-2d_3^{(1)})-\gamma+\ln{2})\right
]\nonumber\\
&+\frac{a}{(\ln{\frac{1}{a}})^3}\Big [-1-\frac{\tilde a_\tau^{(2)}}{2}+2 d_1^{(1)}+(2+\tilde a_\tau^{(0)}) d_2^{(2)}+6 d_3^{(1)}+4\tilde a_\tau^{(0)}d_3^{(1)}-2 \tilde a_\tau^{(1)} d_3^{(1)}\nonumber\\
& \qquad-2 \tilde a_\tau^{(0)} d_3^{(2)}+\frac{\pi ^2}{6}+\frac{24 \tilde a_\tau^{(0)} d_3^{(1)}}{-12+\pi ^2}-(\ln{2})^2+d_2^{(1)} (4+\tilde a_\tau^{(1)}+2 \gamma-2\ln{2})\nonumber\\
& \qquad\qquad \qquad+2\ln{2}+\gamma (-2-\gamma+2\ln{2})\Big ]+O\left (\frac{a}{(\ln{\frac{1}{a}})^4}\right )=0,
\label{c2tot1}\\
& \partial_\tau c_3 +  \frac{a}{\ln{\frac{1}{a}}} \left [2d_3^{(1)}\right ]+\frac{a}{(\ln{\frac{1}{a}})^2}\left[-1+2 (1+\tilde a_\tau^{(0)})  d_3^{(1)}+2 d_3^{(2)}\right ]\nonumber \\
& \qquad \qquad \qquad \qquad \qquad\qquad \qquad \qquad \qquad +O\left (\frac{a}{(\ln{\frac{1}{a}})^3}\right )=0. \label{c3tot1}
\end{eqnarray}
Here we have neglected the expansion coefficients
$c_j$ for $j>3$ by setting $c_4=c_5=c_6=\ldots=0$.  In
Equations~(\ref{c1tot1})-(\ref{c3tot1}) we keep the necessary number of
orders in $\frac{1}{\ln{\frac{1}{a}}}$ to obtain the closed
expressions for the expansion terms in \eqref{atau}.
Equations~(\ref{c1tot1})-(\ref{c3tot1}) can be viewed as the compatibility
conditions which ensure that expansions (\ref{cexp1}) and (\ref{atau})
are correct, so that $a$ is indeed the adiabatically slow variable.

It follows immediately from Eq.~\eqref{c3tot1} in the order
$\frac{a}{\ln{\frac{1}{a}}}$ that
\begin{eqnarray} \label{d3p1value}
d_3^{(1)}=0,
\end{eqnarray}
and from Eq.~\eqref{c2tot1} in the order $\frac{a}{\ln{\frac{1}{a}}}$ that
\begin{eqnarray}\label{atau1}
\tilde a_\tau^{(0)}=-2.
\end{eqnarray}
Then, from Eq.~\eqref{c1tot1} in the order $\frac{a}{\ln{\frac{1}{a}}}$ we obtain
\begin{eqnarray} \label{d1p1value}
d_1^{(1)}=-\frac{1}{2}.
\end{eqnarray}
Using Eqs.~\eqref{c2tot1} and \eqref{d3p1value}-\eqref{d1p1value} we
obtain in the order $\frac{a}{(\ln{\frac{1}{a}})^2}$ that
\begin{eqnarray}\label{atau2}
\tilde a_\tau^{(1)}=-2-2\gamma+2\ln{2}.
\end{eqnarray}
Using Eqs.~\eqref{c1tot1} and \eqref{d3p1value}-\eqref{atau2} we
obtain in the order $\frac{a}{(\ln{\frac{1}{a}})^2}$ that
\begin{eqnarray}\label{dc2p1an}
d_1^{(2)}=\frac{1}{2} (-2+4 d_2^{(1)}-\gamma+\ln{2}).
\end{eqnarray}
Similar, using Eqs.~\eqref{c3tot1} and \eqref{d3p1value}-\eqref{atau2}
we obtain in the order $\frac{a}{(\ln{\frac{1}{a}})^2}$ that
\begin{eqnarray}\label{dc3p2an}
d_3^{(2)}=\frac{1}{2}.
\end{eqnarray}
Equation~\eqref{c2tot1} in order $\frac{a}{(\ln{\frac{1}{a}})^3}$ requires
also to take into account $\partial_\tau c_2$ which is given by
\begin{eqnarray} \label{c2tau}
\partial_\tau c_2=-\frac{1}{2}\frac{a}{(\ln{\frac{1}{a}})^3}+O\left (\frac{a}{(\ln{\frac{1}{a}})^4}\right ),
\end{eqnarray}
according to~\eqref{ctauall} and~\eqref{cexp1}.

Using Eq. (\ref{c2tot1}) in order
$\frac{a}{(\ln{\frac{1}{a}})^3}$ and \eqref{d3p1value}-\eqref{atau2},
\eqref{dc3p2an}, \eqref{c2tau} we obtain the closed expression
\begin{eqnarray}\label{atau3}
\tilde a_\tau^{(2)}=\frac{\pi^2}{3} - 2 (\ln{2})^2 + 4\ln{2} +
 \gamma (-4 - 2\gamma + 4\ln{2}).
\end{eqnarray}
Here, the unknown coefficient $d_2^{(1)}$ has been cancelled out identically.

\begin{figure}
\begin{center}
\includegraphics[width = 3.98 in]{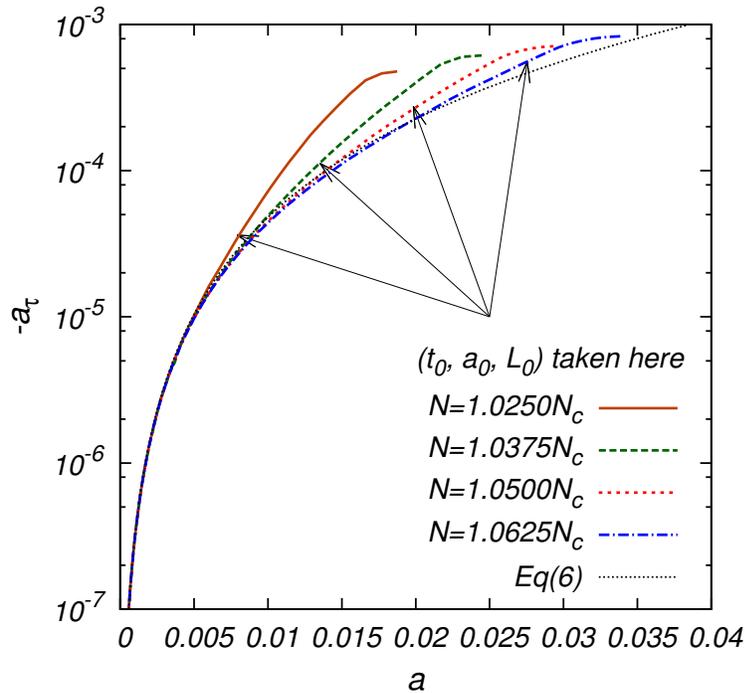}
\end{center}
\caption{Dependence $\partial_\tau a(a)$ extracted from RKSE
  simulations shown in Figure~\ref{fig:Lthirdorder} (thick solid,
  dashed, short-dashed and dashed-dotted lines, respectively).  The thin dotted line represents the
  analytical dependence $\partial_\tau a(a)$ from
  Eq.~\eqref{ataufullnextorder}, with neglected $O(\ldots)$ term.
  Notice that after the initial transient all numerical curves
  collapse on the analytical curve.  The arrows point to the locations
  where the relative difference between the analytical and numerical
  curveds reduces to $20\%$, the criterion for selecting $t_0$ in
  Eq.~\eqref{scaling5} and in Figure~\ref{fig:Lthirdorder}.
}
\label{fig:ataua}
\end{figure}

Equations \eqref{atau}, \eqref{atau1}, \eqref{atau2}, and
(\ref{atau3}) result in closed ODE \eqref{ataufullnextorder} for $a$,
which is the first main result of this paper.  Figure \ref{fig:ataua}
shows $\partial_\tau a$ as a function of $a$ for RKSE simulations
with different initial conditions (the same initial conditions as in
Figure \ref{fig:Lthirdorder}).  Notice that after an initial transient
all curves collapse to the single curve given by
Eq.~\eqref{ataufullnextorder}.  This suggests that we can use the
proximity of numerical curves to the analytical curve as the criterion
for selecting $t_0$ in equation \eqref{scaling5}.  In Figure~\ref{fig:Lthirdorder}, we used the values of $t_0$ defined
for each initial condition as the time $t=t_0$ when the relative difference
between numerical and analytical curves reduces down to $20\%$.  Arrows in
Figure \ref{fig:ataua} point to locations $(a(t_0), \partial_\tau
a(t_0))$ satisfying this criterion. 
For the simulations of Figure \ref{fig:Lthirdorder} we obtained  $t_0 =  7.2125\ldots,$
 $t_0 =  4.5879\ldots,$
 $t_0 =  3.3257\ldots,$
 $t_0 =   2.5528\ldots$ for $N/N_c=1.0250, \, 1.0375, \, 1.0500, \, 1.0625,$ respectively.
 Also in these cases $t_c=  8.12305\ldots, $
$t_c= 5.32533\ldots, $
$t_c=  3.94247\ldots, $
$t_c=  3.12039\ldots,$ respectively.

The dashed-dotted curves in
Figure~\ref{fig:Lthirdorder} are only weakly sensitive to the choice
of $t_0<t_c$, provided $t_0$ is chosen later than the time specified by
the $20\%$-difference criterion.  For instance, if we choose $t_0$
based on $10\%$-difference criterion (instead of  $20\%$), the $L(t)$
curves in Figure~\ref{fig:Lthirdorder} would change by $<5\%$ which is within the relative error of these curves in comparison with the numerics (solid curves in Figure ~\ref{fig:Lthirdorder}).

It also follows from Eqs.~\eqref{cexp1}, \eqref{d3p1value},
\eqref{d1p1value}, and \eqref{dc3p2an} that the expansion
coefficients in~\eqref{vexp1var} are given by the following expressions
\begin{equation} \label{cexp2}
\begin{split}
& c_1=\frac{1}{2}-\frac{1}{2\ln{\frac{1}{a}}}+\frac{1}{2} (-2+4 d_2^{(1)}-\gamma+\ln{2})\frac{1}{(\ln{\frac{1}{a}})^2}+O\left (\frac{1}{(\ln{\frac{1}{a}})^3}\right ),
\\
& c_2=-\frac{1}{2}+\frac{d_2^{(1)}}{\ln{\frac{1}{a}}}+O\left (\frac{1}{(\ln{\frac{1}{a}})^2}\right ),
\\
& c_3=\frac{1}{2(\ln{\frac{1}{a}})^2}+O\left (\frac{1}{(\ln{\frac{1}{a}})^3}\right ).
\end{split}
\end{equation}
Thus, the coefficient $c_3$, which corresponds to positive eigenvalue, is of a lower order compare with $c_1$ and
$c_2.$  We expect the similar to be true for all coefficients $c_3, \,
c_4, \ldots$.  Also, the coefficient $d_2^{(1)}$ is undetermined
in our approximation order.  We expect that it might depend on initial
conditions.  We conclude that the self-similar solution
(\ref{mselfsimilar}) is stable with respect to radially-symmetric
perturbations, and the leading order corrections to it are determined
by a linear combination $v\simeq c_1\psi_1+c_2\psi_2,$ where $c_1$ and
$c_2$ are given by (\ref{cexp2}).

\section{Blow-up rate of self-similar solution }
\label{section:blowuprate}

In this section we solve ODE~\eqref{ataufullnextorder} together
with~\eqref{adef1} and~\eqref{taudef1} to derive the blow-up
rate~\eqref{pselfsimilarfull}.  Integration of
Eq.~\eqref{ataufullnextorder} from an initial value $\tau_0$ to $\tau$
gives
\begin{eqnarray} \label{phieq}
\left . \frac{1}{a}\left [\ln{\frac{1}{a}}+\tilde M+\frac{b}{\ln{\frac{1}{a}}}+O\left (\frac{1}{(\ln{\frac{1}{a}})^2}\right ) \right ]\right|^{\tau}_{\tau=\tau_0}=2(\tau-\tau_0),
\end{eqnarray}
where $\tilde M=\frac{M}{2}-1=-2 - \gamma + \ln{2}$ and $b=\frac{b_0}{2}+\frac{M^2}{4}=1+\frac{\pi^2}{6} $ as
in~\eqref{scaling5}. If we look at Eq.~\eqref{phieq}
as the implicit expression to determine $a(\tau)$ then it turns into a  remote relative of the
Lambert W-function. Such implicit expression can be solved for $a$ assuming $\tau\gg 1$ by
iterations as follows:
\begin{equation} \label{phieqsol}
\ln{\frac{1}{a}}=L_1-L_2+\frac{L_2}{L_1}+\frac{L_2^2}{2L_1^2}-\frac{L_2}{L_1^2}-\frac{\tilde M}{L_1}
+\frac{-2 b+2 \tilde M+\tilde M^2-2 \tilde M L_2}{2 L_1^2}
+O\left (\frac{L_2^3}{L_1^3}\right ),
\end{equation}
where $L_1:= \ln{[2(\tau-\tau^*)]}$, $L_2:= \ln{\ln{[2(\tau-\tau^*)]}}$,  and
\begin{eqnarray} \label{taustar}
\tau^*=\tau_0-\frac{1}{2a}\left (\ln{\frac{1}{a}}+\tilde M+\frac{b}{\ln{\frac{1}{a}}} \right )\Big |_{a=a_0}, \quad a_0=a(\tau_0).
\end{eqnarray}
At this point, one can proceed in qualitatively similar way to
Ref.~\cite{LushnikovPhysLettA2010} to determine $L(\tau)$.  However that way of calculation results in a slow convergence of the asymptotic series for $L(\tau)$ with the increase of $\tau$.   We
choose a different path.  Our goal is to start with Eq.~\eqref{phieq},
to carry as many steps of exact transformations as possible, and to
perform asymptotic expansions as late as possible.
Here and below we abuse notation and use the same notations for all functions
with the same physical meaning, independently of their arguments:
$L=L(t)=L(\tau)=L(a)$, $\tau=\tau(t)=\tau(L)=\tau(a)$ and
$a=a(t)=a(\tau)=a(L) $.  Similar, for initial values
$L_0=L(t_0)=L(\tau_0)=L(a_0)$, $\tau_0=\tau(t_0)=\tau(L_0)=\tau(a_0)$
and $a_0=a(t_0)=a(\tau_0)=a(L_0)$.

We use Eqs. \eqref{adef1} and \eqref{taudef1} to express $a$  through
$\tau$-derivative of $L$ as follows
\begin{equation}\label{adef2}
a = -\frac{\partial_\tau L}{L}.
\end{equation}
We integrate \eqref{adef2} in $\tau$ between $\tau_0$ and $\tau$,
using the integration by parts, to obtain
\begin{align}\label{LnLL0}
&-\ln \dfrac{L}{L_0} =  \int_{\tau_0}^{\tau} a(\tau)\,d\tau = a\tau(a)-a_{0}\tau(a_0) - \int_{a_0}^a \tau \,da \nonumber \\
&\qquad\quad \ \ =\left[\tau - \tau^{*}\right]a - \left[\tau_{0}-\tau^{*}\right]a_0 -\int_{a_0}^{a}(\tau-\tau^{*})\,da.
\end{align}

To evaluate integral over $a$ in~\eqref{LnLL0} explicitly we use $\tau(a)$
from~\eqref{phieq} with \eqref{taustar} and obtain
\begin{align} \label{LnL0x}
 -\ln{\frac{L}{L_0}}  =\dfrac{1}{4}\left [\left(\ln{\frac{1}{a}}\right )^2-\left(\ln{\frac{1}{a_0}}\right )^2\right ]+\dfrac{\tilde M+1}{2}\left({\ln{\frac{1}{a}}}-\ln{\frac{1}{a_0}}\right)\nonumber \\
+\dfrac{b}{2}\left(\ln {\ln{\frac{1}{a}}}-\ln {\ln{\frac{1}{a_0}}}\right) +
\dfrac{b}{2}\left (\dfrac{1}{\ln{\frac{1}{a}}}- \dfrac{1}{\ln{\frac{1}{a_0}}}\right) +O\left ( \frac{1}{\ln{\frac{1}{a}}}\right ).
\end{align}
Note also that the term $O\left ( \frac{1}{\ln{\frac{1}{a}}}\right )$
in~\eqref{LnL0x} originates from the next order term $O\left
(\frac{1}{(\ln{\frac{1}{a}})^2}\right )$ in
Eq. \eqref{phieq}. Formally, in Eq.~\eqref{LnL0x}, the terms $O\left (
\frac{1}{\ln{\frac{1}{a}}}\right )$ and $\frac{b}{2\ln{\frac{1}{a}}}$
are of the same order.  Yet, our numerical simulations indicate that
$\frac{b}{2\ln{\frac{1}{a}}}$ term improves accuracy of the analytic
approximation, so we keep this term in its explicit form.

We introduce new variables,
\begin{equation}\label{ll0def}
l: = \ln{\dfrac{1}{L}} \quad \mbox{and} \quad
l_0: = \ln{\dfrac{1}{L_0}},
\end{equation}
as well as define
\begin{equation}\label{lstardef}
 l^* = l_0 - \dfrac{1}{4}\left(\ln{\frac{1}{a_0}}\right)^2 -
\dfrac{\tilde M+1}{2}\ln{\frac{1}{a_0}}-\dfrac{b}{2}\left (\ln{\ln{\frac{1}{a_0}}}+\dfrac{1}{\ln{\frac{1}{a_0}}}\right),
\end{equation}
which allows to rewrite~\eqref{LnL0x}
as follows:
\begin{equation}\label{llstardef}
l-l^* = \dfrac{1}{4}\left(\ln{\frac{1}{a}}\right)^2 +
\frac{\tilde M+1}{2}\ln{\frac{1}{a}}+\dfrac{b}{2}\left (\ln{\ln{\frac{1}{a}}}+\dfrac{1}{\ln{\frac{1}{a}}}\right )
+O\left ( \frac{1}{\ln{\frac{1}{a}}}\right ).
\end{equation}

We now solve Eq.~\eqref{llstardef} for $\ln{\frac{1}{a}}$.  Instead of
doing straightforward iterations, we neglect the terms
$\frac{b}{2}(\ldots)$, $O\left ( \frac{1}{\ln{\frac{1}{a}}}\right )$
in Eq.~\eqref{llstardef} and solve the remaining part of the equation,
$Y_0^2-2Y_0-V =0$, exactly:
\begin{equation}\label{y0def}
Y_0  = 1 + \sqrt{1 + V},
\end{equation}
where we define
\begin{equation}\label{YVdef}
Y:=-\dfrac{\ln{\frac{1}{a}}}{\tilde M+1} \quad \mbox{and} \quad V:= \dfrac{4}{(\tilde M+1)^2}(l-l^*)
\end{equation}
with $Y_0$ being the leading order approximation to $Y$, such that
\begin{equation}\label{deltaYdef}
Y=Y_0+\delta Y.
\end{equation}

To find $\delta Y$ as a function of $V$, we represent $\delta Y$
through the formal series $\delta Y=\sum_{n=1}^{\infty} \frac{\delta
  Y_{-n}}{Y_0^n}$ (with $Y_0$ given by~\eqref{y0def}).  We use this
series together with \eqref{y0def}-\eqref{deltaYdef} to  perform a series expansion of
Eq.~\eqref{llstardef} in inverse powers of $Y_0$. It allows to determine the coefficients $\delta Y_n$
recursively at integer inverse powers of $Y_0$ starting with the power
zero.  In particular, the zero power gives $ Y_{-1}= -\frac{b\ln\left
  [-(1+\tilde M)Y_0 \right ]}{(1+\tilde M)^2}$.  Note that the double
logarithm $\ln{\ln{\frac{1}{a}}}$ in \eqref{llstardef} also needs to
be expanded. All together it results in
\begin{equation}\label{deltaYexpr}
\delta Y  = -\frac{b\ln\left [-(1+\tilde M)Y_0 \right ]}{(1+\tilde M)^2Y_{0}}+\frac{-b(1+\tilde M)\ln\left [-(1+\tilde M)Y_0 \right ]+b}{(1+\tilde M)^3Y_{0}^2}+O\left (\frac{1}{Y_0^2}\right )+O\left (\frac{\left (\ln Y_0 \right )^2}{Y_0^3}\right ).
\end{equation}
Here, similar to~\eqref{LnL0x}, we keep the term $\frac{b}{(1+\tilde
  M)^3Y_0^2}$, even though this term is of the same order as $O\left
(\frac{1}{Y_0^2}\right )$ term.  Here, $\tilde M+1=-0.884068\ldots$
according to~\eqref{scaling5}. Note, that instead of performing an
expansion in inverse powers of $Y_0$, one can simply do it in inverse
powers of $V^{1/2}$.  This, however, would result in a slower
convergence for moderate ($V\gtrsim 1$) values of $V$.

We rewrite~\eqref{adef1} as $-\frac{L\, dL}{a}=dt$, and integrate it
between time $t_c$ and $t$:
\begin{equation}\label{tctdef}
\int\limits_t^{t_c}dt'=t_c-t=-\int \limits^0_L \frac{L'}{a(L')}dL',
\end{equation}
where following \eqref{ll0def} and \eqref{YVdef} we can represent $L$
through $V$ as $L = \exp{\left (-\left[l^* + \dfrac{(\tilde
      M+1)^2}{4}V\right]\right )}$. The dependence $a(L)$ in
\eqref{tctdef} follows from
\eqref{ll0def}-\eqref{deltaYexpr}. Switching from integration over
$L$ to the integration over $Y_0$ in \eqref{tctdef} we obtain:
\begin{eqnarray}\label{Y0int}
&t_c-t=\int \limits^\infty_{Y_{0}} \exp{\left (-2\left[l^* + \dfrac{(\tilde M+1)^2}{4}[(Y'_0-1)^2-1]\right]\right )}\frac{(\tilde M+1)^2}{2}(Y_0'-1)\nonumber \\
&
\times \exp\left(-(1+\tilde M)Y_0' +\frac{b\ln\left [-(1+\tilde M)Y'_0 \right ]}{(1+\tilde M)Y'_{0}}\right .\nonumber \\
&\left .-\frac{-b(1+\tilde M)\ln{\left [-(1+\tilde M)Y'_0 \right ]}+b}{(1+\tilde M)^2Y_{0}^{'2}}+O\left (\frac{1}{Y_0^{'2}}\right )
\right ) dY_0'.
\end{eqnarray}
Here, the integration cannot be carried explicitly.  Instead, we use the
Laplace method (see
e.g. \cite{SidorovFedoryukShabuninComplexAnalysisBook1985,OlverBook1985})
to evaluate the integral asymptotically in the limit $Y_0\gg 1.$ We
introduce in Eq. \eqref{scaling5} a new integration variable,
\begin{equation}\label{Lambdadef}
z:=Y^{'}_0-Y_0,
\end{equation}
and rewrite Eq. \eqref{Y0int} as
\begin{align}\label{Lambdaint}
t_c-t=\frac{(\tilde M+1)^2}{2}\exp{\left[-2l^*-\frac{(\tilde M+1)^2}{2}Y_0^2+(\tilde M+1)\tilde M Y_0  \right ]}\int \limits^\infty_{0}e^{Y_0 S(z,Y_0)}dz,
\end{align}
where
\begin{align}\label{Sdef}
&S(Y_0,z)=-(\tilde M+1)^2z+\frac{1}{Y_0} \left[-\frac{(\tilde M+1)^2}{2}z^2+(\tilde M+1)\tilde M z+\ln(Y_0+z-1)\right] \nonumber \\
&+\frac{b\ln\left [-(1+\tilde M)(Y_0+z) \right ]}{Y_{0}(1+\tilde M)(Y_0+z)}-\frac{-b(1+\tilde M)\ln{\left [-(1+\tilde M)(Y_0+z) \right ]}+b}{Y_{0}(1+\tilde M)^2(Y_0+z)^{2}}\nonumber \\
&\qquad \qquad \qquad\qquad \qquad \qquad\qquad \qquad \qquad\qquad \qquad +O\left (\frac{1}{Y_{0}(Y_0+z)^2}\right ).
\end{align}
%

To use the Laplace method for asymptotic expansion of the integral in
\eqref{Lambdaint}, we start with the following general expression,
 \cite{SidorovFedoryukShabuninComplexAnalysisBook1985,OlverBook1985}:
\begin{equation}\label{LaplaceExp}
\int \limits^\infty_{0}e^{Y_0 S(z,Y_0)}dz=e^{Y_0 S(0,Y_0)}\displaystyle\sum_{n = 0}^{\infty}c_nY_{0}^{-n-1}
\end{equation}
with
\begin{equation}\label{cndef}
c_n = \left. (-1)^{n+1}\left ( \frac{1}{S^{'}(z,Y_{0})}\frac{\partial}{\partial z}\right )^n
\left
 ( \frac{1}{S^{'}(z,Y_{0})}\right )
 \right |_{z = 0}, \quad S^{'}(z,Y_{0}):=\frac{\partial}{\partial z} S(z,Y_{0}).
\end{equation}
Taking into account two leading terms in \eqref{LaplaceExp}, we obtain
from \eqref{Lambdaint}, \eqref{Sdef}, \eqref{LaplaceExp}, and
\eqref{cndef} the following expression: %
\begin{align}\label{tct0Laplace}
&t_c-t=\frac{(\tilde M+1)^2}{2}\exp{\left[-2l^*-\frac{(\tilde M+1)^2}{2}Y_0^2+(\tilde M+1)\tilde M Y_0 +\ln(Y_0-1) \right ]} \nonumber \\
&\times \exp{\left[\frac{b\ln\left [-(1+\tilde M)Y_0 \right ]}{(1+\tilde M)Y_{0}}-\frac{-b(1+\tilde M)\ln{\left [-(1+\tilde M)Y_0 \right ]}+b}{(1+\tilde M)^2Y_{0}^{2}}+O\left (\frac{1}{Y_0^{2}}\right )
   \right ] }   \nonumber \\
   &\times \frac{1}{(\tilde M+1)^2 Y_0}\left[1+\frac{\tilde M}{(1+\tilde M)Y_{0}}+\frac{\tilde M^2}{(1+\tilde M)^2Y_{0}^{2}}+O\left (\frac{\ln{Y_0}}{Y_0^{3}}\right ) \right ].
\end{align}

We now define a large parameter
\begin{align} \label{xdef2}
x:=\sqrt{-2\ln{\beta (t_c-t)}}-\tilde M,
\end{align}
where
\begin{align} \label{betadef}
\beta:=2\exp{\left \{2 l^* -\frac{\tilde M^2}{2}  \right \}}.
\end{align}

We multiply both the lhs and the rhs of \eqref{tct0Laplace} by $\beta$ from
\eqref{betadef} and take logarithm from both sides to obtain
$-\frac{x^2}{2}$ on the lhs.  We solve the resulting equation for $Y_0$
by assuming the asymptotic form,
\begin{align} \label{Y0xser}
Y_0=b_{-1}x+\sum\limits_{n=0}^\infty \frac{b_n}{x^n}.
\end{align}
and  performing a series expansion of  both rhs and lhs of that resulting equation in inverse powers of $x$.
The coefficients
$b_{-1}, \ ,b_1, \ldots, \ b_3$ are determined recursively giving
\begin{eqnarray} \label{Y0series}
Y_0=-\frac{1}{\tilde M+1}\left [x+\frac{1-b \ln{x}}{x^2}+\frac{-\frac{1}{2}-2\tilde M+b (\ln{x}+2\tilde M\ln{x}-1)}{x^3}\right ]\nonumber \\
\qquad\qquad\qquad\qquad\qquad\qquad\qquad\qquad +O\left (\frac{1}{x^3}\right )+O\left (\frac{(\ln{x})^2}{x^4}\right ).
\end{eqnarray}
Note that the choice of the factor $\exp{\left \{ -\frac{\tilde
    M^2}{2}\right \}}$ in \eqref{betadef} is somewhat arbitrary (the lhs
and the rhs of \eqref{tct0Laplace} can be multiplied by an arbitrary
positive constant). The factor $\exp{\left \{ -\frac{\tilde M^2}{2}\right \}}$
is chosen to speed up convergence of \eqref{Y0series} for $Y_0\gtrsim
1$, i.e. for $L(t)\lesssim 1.$

Using \eqref{ll0def}, \eqref{y0def}, and \eqref{YVdef}
we obtain
\begin{align} \label{LtY0}
L(t)=
\exp{\left [-l^*-\frac{(\tilde M+1)^2}{4}(Y_0^2-2Y_0) \right ]}.
\end{align}
Equations \eqref{ll0def}, \eqref{lstardef} \eqref{xdef2},
\eqref{betadef}, \eqref{Y0series}, and \eqref{LtY0} give the closed
expression for $L(t)$ as a function of  $t_c-t$  and the initial values
$L_0=L(t_0), \ a_0=-L L_t|_{t=t_0}$.  To make the comparison with the
old scaling \eqref{Ltexplog3} more transparent, we plug the
expression for $Y_0$ from \eqref{Y0series} into \eqref{LtY0} and
perform a series expansion of the expression in the exponent into inverse powers of
$x$, obtaining the final expression \eqref{scaling5}. Note that the
first term in the exponent of the first equation in \eqref{scaling5}
can be rewritten through $x$ as $-\sqrt{-\frac{\ln{\beta
      (t_c-t)}}{2}}=-\frac{x+\tilde M}{2}$.  Thus,
Eq.~\eqref{scaling5} includes terms of orders $x$ and $x^0$.

The error terms $O\left (\frac{1}{x^3}\right )$ in~\eqref{Y0series}
and $O\left (\frac{1}{x^2}\right )$ in \eqref{scaling5} result from
the error term $O\left (\frac{1}{(\ln{\frac{1}{a}})^4}\right )$ in
Eq.~\eqref{ataufullnextorder}. We, however, chose to write
down explicitly the terms of the same orders, $\propto\frac{1}{x^3}$ in
\eqref{Y0series} and $\propto \frac{1}{x^2}$ in \eqref{scaling5}. These
terms are independent from the error term $O\left
(\frac{1}{(\ln{\frac{1}{a}})^4}\right )$ of
Eq. \eqref{ataufullnextorder}.  Next order error terms are $O\left
(\frac{(\ln{x})^2}{x^4}\right )$ in~\eqref{Y0series} and $O\left
(\frac{(\ln{x})^2}{x^3}\right )$ in~\eqref{scaling5}.

\section{Numerical simulations of RKSE}
\label{section:simulationalgorithm}

In our numerical simulation we evolve Eq.~\eqref{meq1}, written in
terms of the mass of bacteria $m(r,t)$ within the circle of radius $r$ as defined in
\eqref{mdef}.  The density,
$
\rho(r,t) = \frac{1}{r}\frac{\partial m}{\partial r},
$
and other quantities characterizing the evolution of the collapse are computed
from the mass.  To find the width of the collapse, we assume that the
solution has reached its self-similar form given by
Eq.~\eqref{pselfsimilar}.  Then, the collapse width can be estimated
from the density at the center as $L=(\frac{1}{8}\rho|_{r=0})^{-1/2}$.
To compute the slow parameter, $a$, we differentiate $L(t)$, as
in~\eqref{adef1}.  The self-similar time, $\tau$, is found by
integrating $L(t)$ according to Eq.~\eqref{taudef1}.

A typical solution for $m(r,t)$ is shown in Figure~\ref{fig:rhor}b.
The spatial extent of the collapse is marked by the large gradient of the
solution near the center, which becomes even larger and moves even closer to
the center as time progresses.  This requires special
treatment to ensure that the solution remains well-resolved.

\begin{figure}
\begin{center}
\includegraphics[width = 5.98 in]{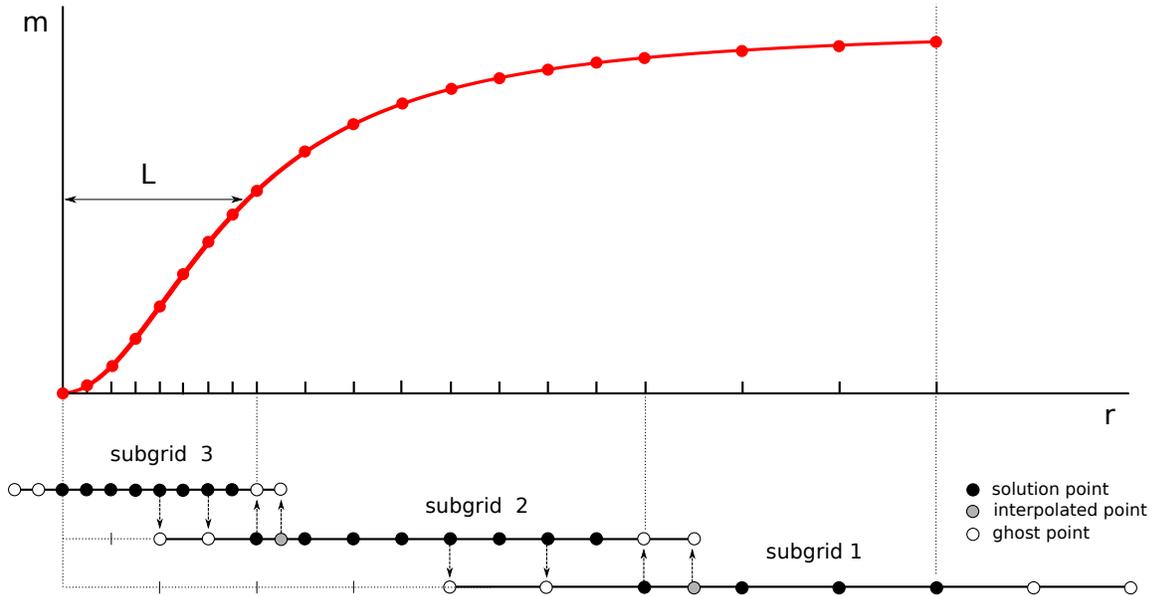}
\caption{Schematic representation of the discretized solution and the
  grid structure.  Three subgrids are shown.  The subgrids closer to
  the center of the collapse have finer resolution.  The data at black
  points are evolved by the discretized Eq.~\eqref{meq1}, the data at white points is copied
  from neighboring subgrids (the copying is shown by arrows), the data
  in gray points is interpolated from neighboring points using
  $6^{th}$ order polynomial.}
\end{center}
\label{fig:amr_grid_example}
\end{figure}

The results presented in this paper are obtained using an adaptive mesh
refinement (AMR) technique~\cite{Berger198964,SulemSulem1999}, complemented with
the  fourth-order Runge-Kutta time advancement method.  Our spatial
domain, $r\in [0,r_{max}]$, is divided into several subdomains
(subgrids) with different spatial resolution. The spacing between
computational points is constant for each subgrid, and differs by a
factor of two between adjacent subgrids.  The rightmost subgrid,
farthest from the collapse, has the coarsest resolution; the spatial
step decreases in the inward direction.

The grid structure adapts during the evolution of the collapse to keep the solution well
resolved.  When a refinement condition is met, the leftmost subgrid is
divided in two equal subgrids.  Then, the new leftmost subgrid is
refined; that is, additional computational points are placed halfway
between the existing points.  The values at the new points are
obtained with sixth-order interpolation.  The condition for grid
refinement comes naturally from properties of the self-similar
profile.   The density at the origin increases by a factor of
$4$ as  the width of profile decreases by a factor of $2$ according to Eq.~(\ref{pselfsimilarfull}).
To keep the effective number of grid points per $L$ within desired
limits, we use the increase of the maximum density by factor of $4$ as the
condition for refinement.

In the interior of each subgrid, the spatial derivatives are
computed using fourh-order central differences on the five point
stencil.
At the subgrid boundaries, the data are copied between subgrids to
fill in values at ``ghost points'', as shown by arrows in
Figure~\ref{fig:amr_grid_example}.  Notice that communication between
subgrids is going in both directions: in AMR terminology, data from
the fine subgrid is restricted to the coarse grid ghost points, and data from the
coarse grid is prolongated to the fine grid ghost points.  The data between points
of the coarser subgrids needed for finer subgrid ghost points are obtained by
sixth order interpolation.  The left ghost points of the leftmost
subgrid are filled using reflective boundary conditions.
The point $r=0$ is treated in a special way because of the singularity in the
rhs of Eq.~\eqref{meq1}.  Expanding $m(r,t)$ in a power series in $r$ at
the origin and using the definition
\eqref{mdef} we obtain that $m(r,t)=\frac{\rho(0,t)r^2}{2}+O(r^4)$. This is also consistent with the series expansion in $r$ of   rhs of Eq.~\eqref{meq1}. Thus in the spatial discretization  we set $m(r=0,t)=0$.
The right ghost points of the rightmost subgrid are filled with the
data from the last point.  We found the right boundary conditions to
be very forgiving, which is not surprising considering that
the mass approaches a constant, as $r^{-3}$, when $r\to\infty$.

The solution on all subgrids is evolved with the same timestep,
$\Delta t = C_{CFL} h^2$, where $h$ is the spatial step of the finest
grid and $C_{CFL}$ is the constant.  We typically used $C_{CFL}=0.4$ but also tested a convergence for smaller values of $C_{CFL}$.

We use two kinds of initial conditions. First kind is the Gaussian,
\[
  m|_{t=0} = A\left( 1 - e^{-(r/\sigma)^2} \right),
   \]
   which implies
\[
  \rho|_{t=0} = \frac{2A}{\sigma^2} e^{-(r/\sigma)^2},\
\]
where $\sigma$ and $A$ are the parameters of the initial condition.
Second kind is the modified stationary solution,  $m|_{t=0} = Am_0(r),$ where $m_0(r)$
is given by Eq.~\eqref{pstaticm0}.
Both types of initial data result in similar dynamics for the same values of $N=2\pi A$.
The simulations presented in this paper were performed for Gaussian initial
conditions, with $\sigma = 1$ and $A = 4.1$, $4.15$, $4.2$, and
$4.25$.  The initial grid was comprised of ten subgrids;  the finest
subgrid had 400 points while all other subgrids had 200 points each.
The size of domain was set to $r_{max} = 1600 L_0$.

We have verified the AMR code against an independently developed,
uniform grid code with an adaptive spatial resolution and an
adaptive time step.  Similar to the AMR code, the uniform-grid code evolved
Eq.~\eqref{meq1} using fourth order Runger-Kutta integration in time.
The spatial derivatives were computed spectrally using the {\tt
  FFTW-3} library \cite{FFTWlink}.  Since Fourier transforms require periodic boundary
conditions, the spatial domain was extended to $r \in [-r_{max},
  r_{max}]$ with sufficiently large $r_{max}$ (about $50 L_0$).  As
in the AMR code, we set the value of the mass to zero at the origin
point to avoid the singularity in the rhs of the equation.  Although
uniform in space, the grid resolution was refined at times when the
maximum density increases by a factor of four.  The new grid had twice
as many points with values computed by spectral interpolation.  We run the
uniform-grid code at $C_{CFL}=0.2$.

Although the uniform-grid code was useful for cross-comparison, it was
significantly less efficient than the AMR code.  Typically, we run the
uniform grid code until the peak density reached $\approx 10^5$, which
required $32,768$ gridpoints with grid resolution
$\frac{L}{8} < h < \frac{L}{4}$.  On the
other hand, in the AMR simulations presented here, the density reached
$\approx10^{17}$ on approximately $12,000$ total gridpoints, with
$\frac{L}{100}<h<\frac{L}{50}$ resolution on the finest subgrid.

\section{Conclusion and Discussion}

\label{section:conclusion}

In conclusion, we studied the collapsing solution of the 2D RKSE,
Eq.~\eqref{pottscontinuousKellerSegelreduced}.  To leading order, the
collapsing solution has the self-similar
form~\eqref{pselfsimilarfull}, characterized by the scaling $L(t)$.  Our
analysis of the dynamics of perturbations about the self-similar form
allowed us to find the time dependence of the width of the collapsing
solution given by the scaling~\eqref{scaling5}.  The analysis of the
perturbations is performed by switching to independent ``blow up''
variables~\eqref{taudef1a}, and an unknown function~\eqref{varphidef}.
In the blow-up variables, the analysis of the dynamics of the collapse
reduces to the analysis of the perturbation about the static
solution~\eqref{mselfsimilar}.  The analysis exploits the slow
evolution of the parameter $a$, defined in~\eqref{adef1}, which
originates from the leading order scaling~$L(t)\propto(t_c-t)^{1/2}$.
After applying the gauge transform~\eqref{vdef1}, we expanded the
general perturbations in eigenfunctions of the self-adjoint
linearization operator $\hat {\mathcal{L}}_a$ about the static
solution and derived the system of amplitude equations.  We solved
these amplitude equations approximately to obtain  ODE~\eqref{ataufullnextorder} for
$a(\tau)$. We solve  Eq.~\eqref{ataufullnextorder} asymptotically in the limit $t\to t_c$, together
with~\eqref{adef1} and~\eqref{taudef1} to obtain the scaling \eqref{scaling5}.

We found that both  ODE~\eqref{ataufullnextorder} for $a(\tau)$ and the scaling~\eqref{scaling5} for $L(t)$  are in excellent
agreement with numerical simulations of RKSE.  We compared
the scaling~\eqref{scaling5} with the previously known
scalings~\eqref{Ltexplog1}-\eqref{Ltexplog3} and showed that
scaling~\eqref{Ltexplog2} is the correct asymptotic limit.  However,
this limit dominates only for unrealistically small values $L\lesssim
10^{-10000}$. In contrast, the scaling~\eqref{scaling5} agrees
well with simulations for a quite moderate  decrease of $L(t)$ compared to the initial
condition. E.g.,  Figure~\ref{fig:Lthirdorderfull}  shows that six-fold decrease of $L$ compare with the initial value
 $L(0)$  is enough to achieve  the relative error $\lesssim 7\%$ between simulations and  the scaling~\eqref{scaling5}.

We now discuss the limitations of the analysis of this paper. The
analysis is exact until we derive the amplitude equation~\eqref{psiLpsi}.  At this point we must resort to approximation,
because we can calculate only a finite number of terms in the
amplitude equation.  We approximate the eigenfunctions
$\psi_j$ of $\hat {\mathcal{L}}_a$ as $\tilde \psi_j$, $j=1,2,\ldots$
through the variational analysis (see Section \ref{section:spectrum}).
Such variational approximation itself does not create any obstacle
because one can, at least in principle, expand the general
perturbation about the static solution in functions $\tilde \psi_j$,
provided they form a complete set on the space $L^2([0,\infty), y^3
  dy)$ which corresponds to the scalar product~\eqref{scalarproduct1}.
However, the variational construction of such functions turns out to
be difficult for $j> 3$ (provided we aim to approximate the
eigenvalues $\lambda_j$ of $\hat {\mathcal{L}}_a$ with high
precision). This work is left for the future.  In this paper we limit
the analysis of the amplitude equations for $\tilde \psi_j$, $j=1,2,3$
by setting $c_j =0$ for $j>3$ in the amplitude equation.  We found
that already $\tilde \psi_3$ gives contribution only to the
coefficient $b_0$
in ODE~\eqref{ataufullnextorder},
i.e. to the highest order
term $\frac{b_0}{(\ln{\frac{1}{a}})^3}$ which we take into account.
The other two lower order terms in rhs
of~\eqref{ataufullnextorder}, $-\frac{2}{\ln{\frac{1}{a}}}
+\frac{M}{(\ln{\frac{1}{a}})^2}$, are fully determined by $\tilde
\psi_1$ and $\tilde \psi_2$.  We expect that taking into account the
nonzero values of $\tilde \psi_j$ and $c_j$ for $j>3$ might modify the
value of $b_0$ in ODE for $a(\tau)$ (and respectively modify $b$
in~\eqref{scaling5}).  Such type of calculation presents a technical
challenge and is left for the future.  A potential route for such
calculation might be the approximation of the eigenfunctions $ \psi_j$
using the matched asymptotic
technique~\cite{DejakLushnikovOvchinnikovSigalPhysD2012}.  Using this
technique, however, one faces the challenge of calculating the scalar
products in the amplitude equation.  It should be mentioned that the
dramatic improvement of the accuracy of $L(t)$ with the increase of the order of
approximation (as seen in comparison of Figures~\ref{fig:Lthirdorder}b,c with
Figure~\ref{fig:Lthirdorder}a)
suggests that the essential part of  $b_0$ (and
respectively $b$) is already captured in our scaling~\eqref{scaling5}.


\appendix

\section{Keller-Segel model of bacterial aggregation}
\label{appendix:KellerSegelModel}

Bacteria and biological cells often communicate through chemotaxis, the process of secretion and
detection of a substance called chemoattractant. Below
we refer to bacteria and cell as synonyms. The chemoattractant
secreted by bacteria diffuses through media. Other bacteria of the
same kind detect it and move along its gradient.  Thus the chemotaxis
creates nonlocal attraction between bacteria.  Bacteria are
self-propelled and, without the chemotactic clue, the center of
mass of each bacterium typically experiences a random walk.
The motion of bacterial colonies is thus affected by the competition
between random-walk-based diffusion and chemotaxis-based attraction.
The macroscopically averaged motion of bacteria can be described by
the Keller-Segel model (sometimes called the Patlak-Keller-Segel
equation), see e.g.,
\cite{Patlak1953,KellerSegel1970,Alt1980,HerreroVelazquezMathAnn1996,BrennerLevitovBudrene1998,BrennerConstantinKadanoff1999,Ben-JacobAdvPhys2000,BettertonBrennerPRE2001,VelazquezSIAMJApplMath2002,HillenOthmer2002,SireChavanis2002,ErbanOthmer2005,NewmanGrima2004,LushnikovPhysLettA2010,DejakLushnikovOvchinnikovSigalPhysD2012,DLS2009}
and references therein:
\begin{eqnarray}\label{pottscontinuousKellerSegel}
\partial _t \rho &=&D\nabla^2 \rho-\nabla\big [k \rho\, \nabla c\big ], \\
\partial_tc&=&D_c\nabla^2 c +\alpha\, \rho,  \label{pottscontinuousKellerSegelc}
\end{eqnarray}
where $\rho({\bf r},t)$ is the bacterial density at spatial point
${\bf r}$ and time $t$, $c({\bf r},t)$ is the concentration of
chemoattractant, $D$ is the diffusion coefficient of bacteria
(representing the random walk), $D_c$ is the diffusion coefficient of
chemoattractant, $\alpha$ is the production rate of chemoattractant by
bacteria, and the coefficient $k>0$ characterizes the strength of
chemotaxis.

The Keller-Segel model is a mean-field approximation of the behavior
of a large number of bacteria, and can be derived from the dynamics of
individual bacteria using macroscopic averaging over an ensemble of
realizations of stochastic bacteria motion.  A starting point of the
derivation can be, e.g., the description of an ensemble of bacteria as
point-wise objects subject to a white noise force, as in
Ref.~\cite{NewmanGrima2004}. Such description is most relevant to
procaryotic bacteria like {\it Escherichia coli} which small rigid shapes.  Another possible
starting point is the description of the dynamics of eukaryotic
organisms with randomly fluctuating shape, such as {\it Dictyostelium}
amoeba~\cite{AlberChenGlimmLushnikov2006,AlberChenLushnikovNewman2007,LushnikovChenalberPRE2008}.

If the initial density of bacteria is low, the bacterial diffusion
typically dominates attraction and the density remains low.  For
instance, a typical time scale for the evolution of a low-density {\it
  Escherichia coli} distribution in a petri dish is about one
day~\cite{BrennerLevitovBudrene1998} (see Figure 3A in
Ref.~\cite{BrennerLevitovBudrene1998}).  If the initial density is
relatively high, attraction dominates, and bacteria aggregate (see
Figure 3B in Ref.~\cite{BrennerLevitovBudrene1998}).  The typical time
scale of such aggregation in experiments on {\it Escherichia coli} is
several minutes~\cite{BrennerLevitovBudrene1998}. Thus the aggregation has
an explosive character compared to the
evolution of bacteria outside the aggregation area.  The aggregation is described by the
``collapse of bacterial density'' in the approximation of the Keller-Segel
model~(\ref{pottscontinuousKellerSegel})-(\ref{pottscontinuousKellerSegelc}).

The diffusion of chemoattractant is usually much faster than the
diffusion of bacteria, i.e., $D/D_c\ll 1$.  For instance, $D/D_c\sim
1/40-1/400$ for the cellular slime mold {\it
  Dictyostelium}~\cite{HoferSherrattMainiPhysD1995}, and $D/D_c\sim
1/30$ for microglia cells and
neutrophils~\cite{LucaChavez-RossEdelstein-KeshetMogilner2003,GrimaPRL2005}.
(Here, we refer to bacteria and cell as synonyms.)
Thus Eq.~(\ref{pottscontinuousKellerSegelc}) evolves on a much
smaller time scale than Eq.~(\ref{pottscontinuousKellerSegel}), so we
can neglect the time derivative
in~\eqref{pottscontinuousKellerSegelc}.  In addition, we assume that
$D$, $D_c$, $\alpha$, and $k$ are constants, and recast all variables
in dimensionless form: $t\to t_0 t$, $r\to t_0^{1/2}D^{1/2}r$,
$\rho\to (D_c/t_0\alpha k)\rho$, and $c\to (D/k)c$, where $t_0$ is a
typical timescale of the dynamics of $\rho$ in
Eq.~\eqref{pottscontinuousKellerSegel}.  The resulting system is
called the reduced Keller-Segel
equations~\eqref{pottscontinuousKellerSegelreduced}.

\section{Calculation of scalar products through Meijer G-function and $\Gamma$-function}
\label{appendix:MeijerGfunction}

Calculation of scalar products in Section
\ref{section:amplitudeexpansion} requires to evaluate the integrals of
the following type
\begin{equation} \label{Imnldef}
I^{n,m}_l:=\int\limits_0^\infty \frac{e^{-\frac{a}{2}y^2}y^{2n+1}[\ln{(1+y^2)}]^m}{(1+y^2)^l}dy,  \quad n,m,l \in \mathbb{N}, \quad n\ge0,l\ge2, m\ge 0,
\end{equation}
which by the change of variable $x:=1+y^2$ and differentiation over the parameter $a$ reduces to the following expression%
\begin{eqnarray} \label{Imnldef2}
I^{n,m}_l=(-1)^n 2^{n-1}\frac{d^n}{da^n}\int\limits_1^\infty \frac{e^{-\frac{a}{2}(x-1)}[\ln{x}]^m}{x^l}dx\nonumber \\
\qquad\qquad\qquad=(-1)^n 2^{n-1}m!\frac{d^n}{da^n}\left[
e^{\frac{a}{2}} G^{m+2,0}_{m+1,m+2}\left (\begin{smallmatrix}{\frac{a}{2}}\end{smallmatrix} \Big \vert \begin{smallmatrix}l,\ldots, l \\ 0,l-1,\ldots, l-1\end{smallmatrix} \right ) \right ],
\end{eqnarray}
where $G^{l,k}_{p,q}\left (z \Big \vert \begin{smallmatrix}a_1,\ldots, a_p \\ b_1,\ldots, b_q\end{smallmatrix} \right )$ is the Meijer $G$-function \cite{GradshteynRyzhik2007,PrudnikovMarichevBrychkovVol3p1990}.

E.g., for $n=0$ and $l=2$:
\begin{equation}\label{I012ldef}
\begin{split}
&I^{0,1}_2=\frac{1}{2}e^{\frac{a}{2}} G^{3,0}_{2,3}\left (\begin{smallmatrix}{\frac{a}{2}}\end{smallmatrix} \Big \vert \begin{smallmatrix}2,2 \\ 0,1,1\end{smallmatrix} \right ),\\
&I^{0,2}_2=e^{\frac{a}{2}} G^{4,0}_{3,4}\left (\begin{smallmatrix}{\frac{a}{2}}\end{smallmatrix} \Big \vert \begin{smallmatrix}2,2,2 \\ 0,1,1,1\end{smallmatrix} \right ),\\
&I^{0,3}_2=3e^{\frac{a}{2}} G^{5,0}_{4,5}\left (\begin{smallmatrix}{\frac{a}{2}}\end{smallmatrix} \Big \vert \begin{smallmatrix}2,2,2,2 \\ 0,1,1,1,1\end{smallmatrix} \right ),\\
&I^{0,4}_2=12e^{\frac{a}{2}} G^{6,0}_{5,6}\left (\begin{smallmatrix}{\frac{a}{2}}\end{smallmatrix} \Big \vert \begin{smallmatrix}2,2,2,2,2 \\ 0,1,1,1,1,1\end{smallmatrix} \right ).
\end{split}
\end{equation}
And more generally, for $n=0$ and $l\ge 2$:
\begin{equation}\label{I012ladef}
\begin{split}
&I^{0,1}_l=\frac{1}{2}e^{\frac{a}{2}} G^{3,0}_{2,3}\left (\begin{smallmatrix}{\frac{a}{2}}\end{smallmatrix} \Big \vert \begin{smallmatrix}l,l \\ 0,l-1,l-1\end{smallmatrix} \right ),\\
&I^{0,2}_l=e^{\frac{a}{2}} G^{4,0}_{3,4}\left (\begin{smallmatrix}{\frac{a}{2}}\end{smallmatrix} \Big \vert \begin{smallmatrix}l,l,l \\ 0,l-1,l-1,l-1\end{smallmatrix} \right ),\\
&I^{0,3}_l=3e^{\frac{a}{2}} G^{5,0}_{4,5}\left (\begin{smallmatrix}{\frac{a}{2}}\end{smallmatrix} \Big \vert \begin{smallmatrix}l,l,l,l \\ 0,l-1,l-1,l-1,l-1\end{smallmatrix} \right ),\\
&I^{0,4}_l=12e^{\frac{a}{2}} G^{6,0}_{5,6}\left (\begin{smallmatrix}{\frac{a}{2}}\end{smallmatrix} \Big \vert \begin{smallmatrix}l,l,l,l,l \\ 0,l-1,l-1,l-1,l-1,l-1\end{smallmatrix} \right ).
\end{split}
\end{equation}

A particular case $m=0$ is especially easy because $G$-function from \eqref{Imnldef} reduces to the incomplete Gamma function $\Gamma(s,z)=\int\limits_z^\infty t^{s-1}e^{-t}dt$ as follows
\begin{equation} \label{ImnlGamma}
I^{n,0}_l=(-1)^n 2^{n-l}a^{l-1}\frac{d^n}{da^n}\left [e^{\frac{a}{2}} \Gamma{(1-l,{\frac{a}{2}} )}\right ].
\end{equation}

A Taylor series expansion of \eqref{ImnlGamma} for $a\to 0$ gives for $n=0$ the following expressions
\begin{equation}\label{Inmlasymp3}
\begin{split}
&I^{0,0}_2=\frac{1}{2}+\frac{1}{4} (\gamma-\ln{2}+\ln{a}) a+\frac{1}{8} (-1+\gamma-\ln{2}+\ln{a}) a^2+O(a^3\ln{a})\\
&I^{0,0}_3=\frac{1}{4}-\frac{a}{8}+\frac{1}{16} (-\gamma+\ln{2}-\ln{a}) a^2+\frac{1}{32} (1-\gamma+\ln{2}-\ln{a}) a^3+O(a^4\ln{a}),\\
&I^{0,0}_4=\frac{1}{6}-\frac{a}{24}+\frac{a^2}{48}+\frac{1}{96} (\gamma-\ln{2}+\ln{a}) a^3+\frac{1}{192} (-1+\gamma-\ln{2}+\ln{a}) a^4\\&+O(a^5\ln{a}),\\
&I^{0,0}_5=\frac{1}{8}-\frac{a}{48}+\frac{a^2}{192}-\frac{a^3}{384}+\frac{1}{768} (-\gamma+\ln{2}-\ln{a}) a^4+\frac{(1-\gamma+\ln{2}-\ln{a}) a^5}{1536}\\&\qquad\qquad\qquad\qquad+O(a^6\ln{a}).\\ \end{split}
\end{equation}
and the case $n>0$ is obtained by the differentiation of these  expressions according to \eqref{ImnlGamma}.

A Taylor series expansion of \eqref{Imnldef2} for $a\to 0$ gives for $n=0$ the following expressions for $l=1:$
\begin{equation}\label{Inmlasymp4}
\begin{split}
&I^{0,1}_2=\frac{1}{2}+\frac{1}{48} \Big(12 \gamma-6 \gamma^2-\pi ^2-12 \ln{2}+12 \gamma \ln{2}-6 (\ln{2})^2\\
&+12 \ln{a}-12 \gamma \ln{a}+12 \ln{2} \ln{a}-6 (\ln{a})^2\Big) a+\frac{1}{96} \Big(12 \gamma-6 \gamma^2-\pi ^2-12 \ln{2}+12 \gamma \ln{2}\\&-6 (\ln{2})^2+12 \ln{a}-12 \gamma \ln{a}+12 \ln{2} \ln{a}-6 (\ln{a})^2\Big) a^2+O\left(a^3 (\ln{a})^2\right),\\
&I^{0,1}_3=\frac{1}{8}-\frac{3 a}{16}+\frac{1}{192} \Big(-18 \gamma+6 \gamma^2+\pi ^2+18 \ln{2}-12 \gamma \ln{2}+6 (\ln{2})^2-18 \ln{a}+12 \gamma \ln{a}\\&-12 \ln{2} \ln{a}+6 (\ln{a})^2\Big) a^2+\frac{1}{384} \Big(6-18 \gamma+6 \gamma^2+\pi ^2+18 \ln{2}-12 \gamma \ln{2}+6 (\ln{2})^2\\&-18 \ln{a}+12 \gamma \ln{a}-12 \ln{2} \ln{a}+6 (\ln{a})^2\Big) a^3+O\left(a^4 (\ln{a})^2\right),\\
&I^{0,1}_4=\frac{1}{18}-\frac{5 a}{144}+\frac{11 a^2}{288}+\frac{1}{1152}\Big(22 \gamma-6 \gamma^2-\pi ^2-22 \ln{2}+12 \gamma \ln{2}-6 (\ln{2})^2+22 \ln{a}\\&-12 \gamma \ln{a}+12 \ln{2} \ln{a}-6 (\ln{a})^2\Big) a^3
+O\left(a^4 (\ln{a})^2\right ),\\
&I^{0,1}_5=\frac{1}{32}-\frac{7 a}{576}+\frac{13 a^2}{2304}-\frac{25 a^3}{4608}\\&+\frac{1}{9216}\Big(-25 \gamma+6 \gamma^2+\pi ^2+25 \ln{2}-12 \gamma \ln{2}+6 (\ln{2})^2-25 \ln{a}+\\&12 \gamma \ln{a}-12 \ln{2} \ln{a}+6 (\ln{a})^2\Big) a^4+O\left(a^5 (\ln{a})^2\right),
%
\end{split}
\end{equation}
for $l=2$:
\begin{equation}\label{Inmlasymp4a}
\begin{split}
&I^{0,2}_2=1+\Big[-\frac{1}{24} \left[-12+12 \gamma-6 \gamma^2-\pi ^2-12 \ln{2}+12 \gamma \ln{2}-6 (\ln{2})^2\right] \ln{a}\\&+\frac{1}{4} (-1+\gamma-\ln{2}) (\ln{a})^2+\frac{(\ln{a})^3}{12}\Big]a+O(a)+O\left (a^2 (\ln{a})^3\right),\\
&I^{0,2}_3=\frac{1}{8}-\frac{7 a}{16}+\Big[-\frac{1}{96} \left[21-18 \gamma+6 \gamma^2+\pi ^2+18 \ln{2}-12 \gamma \ln{2}+6 (\ln{2})^2\right] \ln{a}\\&+\frac{1}{32} (3-2 \gamma+2 \ln{2}) (\ln{a})^2-\frac{(\ln{a})^3}{48}\Big]a^2+O(a^2)+O\left (a^3 (\ln{a})^3\right),\\
&I^{0,2}_4=\frac{1}{27}-\frac{19 a}{432}+\frac{85 a^2}{864}\\&+\Big[-\frac{\left(-85+66 \gamma-18 \gamma^2-3 \pi ^2-66 \ln{2}+36 \gamma \ln{2}-18 (\ln{2})^2\right) \ln{a}}{1728}\\&+\frac{1}{576} (-11+6 \gamma-6 \ln{2}) (\ln{a})^2+\frac{(\ln{a})^3}{288}\Big]a^3+O(a^3)+O\left (a^4 (\ln{a})^3\right),\\
&I^{0,2}_5=\frac{1}{64}-\frac{37 a}{3456}+\frac{115 a^2}{13824}-\frac{415 a^3}{27648}+\\&
 \Big[-\frac{\left(415-300 \gamma+72 \gamma^2+12 \pi ^2+300 \ln{2}-144 \gamma \ln{2}+72 (\ln{2})^2\right) \ln{a}}{55296}\\&+\frac{(25-12 \gamma+12 \ln{2}) (\ln{a})^2}{9216}-\frac{(\ln{a})^3}{2304}\Big]a^4+O(a^4)+O\left (a^5 (\ln{a})^3\right),\\
\end{split}
\end{equation}
and for $l=3:$
\begin{equation}\label{Inmlasymp5}
\begin{array}{rlr}
&I^{0,3}_2=3+\Big [\frac{1}{16} \left(-12+12 \gamma-6 \gamma^2-\pi ^2-12 \ln{2}+12 \gamma \ln{2}-6( \ln{2})^2\right) (\ln{a})^2\\
&-\frac{1}{4} (-1+\gamma-\ln{2}) (\ln{a})^3-\frac{(\ln{a})^4}{16}\Big ]a+O(a\ln{a})+O\left (a^2 (\ln{a})^4\right),\\
&I^{0,3}_3=\frac{3}{16}-\frac{45 a}{32}+\Big[\frac{1}{64} \left(21-18 \gamma+6 \gamma^2+\pi ^2+18 \ln{2}-12 \gamma \ln{2}+6 \ln{2}^2\right) (\ln{a})^2\\
&-\frac{1}{32} (3-2 \gamma+2 \ln{2}) (\ln{a})^3+\frac{(\ln{a})^4}{64}\Big]a^2+O(a^2\ln{a})+O\left (a^3 (\ln{a})^4\right),\\
&I^{0,3}_4=\frac{1}{27}-\frac{65 a}{864}+\frac{575 a^2}{1728}\\
&+\Big [\frac{\left(-85+66 \gamma-18 \gamma^2-3 \pi ^2-66 \ln{2}+36 \gamma \ln{2}-18 \ln{2}^2\right) (\ln{a})^2}{1152}
\\
&-\frac{1}{576} (-11+6 \gamma-6 \ln{2}) (\ln{a})^3-\frac{(\ln{a})^4}{384}\Big ]a^3+O(a^3\ln{a})+O\left (a^4 (\ln{a})^4\right),\\
&I^{0,3}_5=\frac{3}{256}-\frac{175 a}{13824}+\frac{865 a^2}{55296}-\frac{5845 a^3}{110592}\\
&+\Big [\frac{\left(415-300 \gamma+72 \gamma^2+12 \pi ^2+300 \ln{2}-144 \gamma \ln{2}+72 \ln{2}^2\right) (\ln{a})^2}{36864}\\
&-\frac{(25-12 \gamma+12 \ln{2}) (\ln{a})^3}{9216}+\frac{(\ln{a})^4}{3072}\Big ]a^4+O(a^4\ln{a})+O\left (a^5 (\ln{a})^4\right).
\end{array}
\end{equation}
The case $n>0$ is obtained by the differentiation of these  expressions according to \eqref{Imnldef2}.

The authors thank I.M. Sigal for many  helpful discussions.


Work of P.L., S.D. and N.V.  was partially supported by NSF grants DMS 0719895 and
DMS 0807131.



\end{document}